\documentclass[a4paper,fleqn,usenatbib,twocolumn]{mnras}

\usepackage[T1]{fontenc}
\usepackage{ae,aecompl}

\usepackage{graphicx}	
\usepackage{amsmath}	
\usepackage{amssymb}	
\usepackage{bm}

\newcommand{\const}{\mathop{\text{const}}\nolimits}
\newcommand{\cotan}{\mathop{\text{cotan}}\nolimits}

\title[Homogeneously derived transit timings for 17 exoplanets]{Homogeneously derived transit timings for 17
exoplanets and reassessed TTV trends for WASP-12 and WASP-4}

\author[Baluev et~al.]{%
R.V.~Baluev$^{1,2}$\thanks{E-mail: r.baluev@spbu.ru}, E.N.~Sokov$^{2,1}$, H.R.A.~Jones$^3$, V.Sh.~Shaidulin$^1$, I.A.~Sokova$^{2,1}$,
\newauthor
L.D.~Nielsen$^4$, P.~Benni$^5$, E.M.~Schneiter$^6$, C.~Villarreal~D'Angelo$^7$,
\newauthor
E.~Fern\'andez-Laj\'us$^{8,9}$, R.P. Di Sisto$^{8,9}$, \"O.~Ba\c{s}t\"urk$^{10}$, M.~Bretton$^{11}$, A.~Wunsche$^{11}$,
\newauthor
V.-P.~Hentunen$^{12}$, S.~Shadick$^{13}$, Y.~Jongen$^{14}$, W.~Kang$^{15}$, T.~Kim$^{15,16}$,
\newauthor
E.~Pak\v{s}tien\.e$^{17}$, J.K.T.~Qvam$^{18}$, C.R.~Knight$^{19}$, P.~Guerra$^{20}$, A.~Marchini$^{21}$,
\newauthor
F.~Salvaggio$^{21}$, R.~Papini$^{21}$, P.~Evans$^{22}$, M.~Salisbury$^{23}$, F.~Garcia$^{24}$, D.~Molina$^{25}$,
\newauthor
J.~Garlitz$^{26}$, N.~Esseiva$^{27}$, Y.~Ogmen$^{28}$, Yu.~Karavaev$^{29}$, S.~Rusov$^2$,
\newauthor
M.A.~Ibrahimov$^{30}$, and R.G.~Karimov$^{31}$}

\date{
      Received 2019 August 13;
      in original form 2019 May 31}

\pubyear{2019}

\begin{document}
\label{firstpage}
\pagerange{\pageref{firstpage}--\pageref{lastpage}}
\maketitle

\begin{abstract}
We homogeneously analyse $\sim 3.2\times 10^5$ photometric measurements for $\sim 1100$
transit lightcurves belonging to $17$ exoplanet hosts. The photometric data cover $16$
years 2004--2019 and include amateur and professional observations. Old archival
lightcurves were reprocessed using up-to-date exoplanetary parameters and empirically
debiased limb-darkening models. We also derive self-consistent transit and radial-velocity
fits for $13$ targets. We confirm the nonlinear TTV trend in the WASP-12 data at a high
significance, and with a consistent magnitude. However, Doppler data reveal hints of a
radial acceleration about $(-7.5\pm 2.2)$~m/s/yr, indicating the presence of unseen distant
companions, and suggesting that roughly $10$ per cent of the observed TTV was induced via
the light-travel (or Roemer) effect. For WASP-4, a similar TTV trend suspected after the
recent TESS observations appears controversial and model-dependent. It is not supported by
our homogeneus TTV sample, including $10$ ground-based EXPANSION lightcurves obtained in
2018 simultaneously with TESS. Even if the TTV trend itself does exist in WASP-4, its
magnitude and tidal nature are uncertain. Doppler data cannot entirely rule out the Roemer
effect induced by possible distant companions.
\end{abstract}


\section*{Authors' affiliations}
{\small
$^1$Saint Petersburg State University, Faculty of Mathematics \& Mechanics, Universitetskij pr. 28, Petrodvorets, St Petersburg 198504, Russia\\
$^2$Central Astronomical Observatory at Pulkovo of Russian Academy of Sciences, Pulkovskoje sh. 65/1, St Petersburg 196140, Russia\\
$^3$University of Hertfordshire, Centre for Astrophysics Research, STRI, College Lane, Hatfield AL10 9AB, UK\\
$^4$Geneva Observatory, University of Geneva, Chemin des Mailettes 51, 1290 Versoix, Switzerland\\
$^5$Acton Sky Portal (Private Observatory), Acton, MA, USA\\
$^6$Instituto de Astronom\'ia Teor\'ica y Experimental, Universidad Nacional de C\'ordoba, Laprida 854, C\'ordoba X5000BGR, Argentina\\
$^7$School of Physics, Trinity College Dublin, The University of Dublin, Dublin 2, Ireland\\
$^8$Facultad de Ciencias Astron\'omicas y Geof\'isicas - Universidad Nacional de La Plata, Paseo del Bosque S/N - 1900 La Plata,\\ Argentina\\
$^9$Instituto de Astrof\'isica de La Plata (CCT La Plata - CONICET/UNLP), Argentina\\
$^{10}$Ankara University, Faculty of Science, Department of Astronomy and Space Science, TR-06100, Tandogan, Ankara, Turkey\\
$^{11}$Baronnies Proven\c{c}ales Observatory, Hautes Alpes - Parc Naturel R\'egional des Baronnies Proven\c{c}ales, F-05150 Moydans, France\\
$^{12}$Taurus Hill Observatory, Warkauden Kassiopeia ry., H\"ark\"am\"aentie 88, 79480 Kangaslampi, Finland\\
$^{13}$Physics and Engineering Physics Department, University of Saskatchewan, 116 Science Place, Saskatoon, Saskatchewan,\\ S7N 5E2, Canada\\
$^{14}$Observatoire de Vaison la Romaine, 1075 RD 51, Le Palis, 84110 Vaison-la-Romaine, France\\
$^{15}$National Youth Space Center, Goheung, Jeollanam-do, 59567, S.Korea\\
$^{16}$Department of Astronomy and Space Science, Chungbuk National University, Cheongju-City, 28644, S.Korea\\
$^{17}$Institute of Theoretical Physics and Astronomy, Vilnius University, Sauletekio al. 3, Vilnius 10257, Lithuania\\
$^{18}$Horten Videreg\r{a}ende Skole, Bekkegata 2, 3181 Horten, Norway\\
$^{19}$Ngileah Observatory, 144 Kilkern Road, RD 1. Bulls 4894, New Zealand\\
$^{20}$Observatori Astron\`omic Albany\`a, Cam\'i de Bassegoda s/n, 17733 Albany\`a, Spain\\
$^{21}$Astronomical Observatory, DSFTA - University of Siena, Via Roma 56, 53100 - Siena, Italy\\
$^{22}$El Sauce Observatory, Coquimbo Province, Chile\\
$^{23}$School of Physical Sciences, The Open University, Milton Keynes, MK7 6AA, UK\\
$^{24}$La Vara, Valdes Observatory, 33784 Munas de Arriba, Valdes, Asturias, Spain\\
$^{25}$Anunaki Observatory, Calle de los Llanos, 28410 Manzanares el Real, Spain\\
$^{26}$AAVSO, Private Observatory, Elgin, OR 97827, USA\\
$^{27}$Observatory Saint Martin, code k27, Amathay Vesigneux, France\\
$^{28}$Green Island Observatory, Code B34, Gecitkale, Famagusta, North Cyprus\\
$^{29}$Institute of Solar-Terrestrial Physics (ISTP), Russian Academy of Sciences (Siberian Branch), Irkutsk 664033, p.b. 291,\\
Lermontov street 126a, Russia\\
$^{30}$Institute of Astronomy of Russian Academy of Sciences, Pyatnitskaya Str. 48, Moscow
119017, Russia\\
$^{31}$Ulugh Beg Astronomical Institute of Uzbek Academy of  Sciences, Astronomicheskaya
Str. 33, Tashkent 100052, Uzbekistan}


\section{Introduction}
Transit photometry is now one of the primary exoplanets detection tools. This method has a
very promising descedant branch~--- transit timing variations, or TTVs. The outstanding
value of the TTV method comes from its ability to directly detect observable hints of
$N$-body interactions in a planetary system. This method is even capable of detecting
previously unknown planets \citep{TTV}, and directly reveal tidal interactions with the
star, like now famous example of WASP-12~\emph{b} \citep{Maciejewski18a,BaileyGoodman19}.
This star demonstrates subtle period drift, as if the planet was spiraling down onto its
host star. Such a physical phenomenon brings us unique opportunities to test the theories
of tidal planet-star interaction and even to put some constraints on the interior structure
of this exoplanet \citep{Patra17}. Recently, hints of an analogous TTV drift were also
reported for WASP-4 \citep{Bouma19}, based on the first TESS observations.

Our present work is devoted to further development of the TTV method. Basically, it
presents results of a revised analysis following \citep{Baluev15a} but including additional
targets, expanded photometric data, and improved processing algorithms. However, if the
goal of \citet{Baluev15a} was to demonstrate the potential of amateur observations in the
TTV field, the primary accent here is to highlight the importance of using homogeneously
derived TTVs.

The exoplanetary transit times published in literature are derived by multiple independent
teams that used very different methods and models. For example, some works assume linear
limb-darkening law, but some quadratic. The limb-darkening coefficients may be fixed at
theoretically predicted values, or fitted as free parameters of the lightcurve. The
photometric noise can be modelled differently as well: while some early measurements did
not yet take into account the red noise, others did, but all in different ways. Some tried
to reduce systematic effects by decorrelating them with airmass, some use more complicated
correlation models, and some just fit the systematics by a deterministic model (e.g. trends
plus multiple oscillations).

Moreover, any transit lightcurve fit also depends on the exoplanetary parameters
(planet/star radii ratio, impact parameter, etc.) which have an obvious tendency to improve
their accuracy with time. While many earlier transit observations were rather inaccurate
because they could not rely on good exoplanetary parameters, later ones can use a larger
record of observations to derive more accurate results.

As such, the transit times published in the literature appear very heterogeneous: they may
have subtle systematic biases, including biases in their uncertainties. Those biases are
difficult to deal with, because they vary from one team to another in an impredictable
manner. Hence, it might appear too difficult to analyse such merged TTV data as if they
were homogeneous, ultimately resulting in false detections of spurious variations and so
on.

This work presents an attempt to carefully reprocess the archival and new observations in a
homogeneous way, relying on the same analysis protocol, including the use of the same
methods and of the same transit and noise models. Now we can reprocess the entire
photometry set available for each target in a self-consistent manner, i.e. we should not
necessarily fit all the transits for the same target independently. This approach was
already tested in \citep{Baluev15a}, and it allows us to reduce the number of degrees of
freedom of the fit, thus improving the usability of lower-quality observations.

Such a goal naturally implies substantial analysis of the available photometric data,
careful identification of possibly outlying measurements or even entire lightcurves. Such a
work necessarily implies an investigation of the models involved, in particular the
limb-darkening models and noise models. This also includes an analysis of the photometric
noise potentially yielding improved data-processing strategies.

Moreover, we now aim to undertake a multimethod study not relying on just the photometric
observations. We performed a self-consistent analysis of our homogeneously processed
photometry jointly with Doppler data, since the combination of the transit and Doppler
methods allows for a much more comprehensive characterization of a planetary system. This
is especially important for several unique exoplanets, like the above-mentioned WASP-12 or
WASP-4 demonstrating possible TTV trends. In particular, relatively little attention was
paid so far to a yet another explanation of such trends, based on the light-travel effect
induced by outer bodies \citep{Irwin52}.

Finally, this work represents the first big practical test of the EXPANSION project
(EXoPlanetary trANsit Search with an International Observational Network), grown on the
basis of the ETD (Exoplanet Transit Database) that was used by \citet{Baluev15a}. Now
EXPANSION is a standalone international project joining a network of several dozens of
relatively small-aperture telescopes, aimed to monitor the exoplanetary transits
\citep{Sokov18}. This network covers amateur as well as professional observatories spreaded
over the world in the both hemispheres.

The structure of the paper is as follows. In Sect.~\ref{sec_src} we provide a detailed
description of the data that we analyse. In Sect.~\ref{sec_ttvmethod} we introduce the
algorithms used to process the photometric data. In Sect.~\ref{sec_ld} we present results
of empirical debiasing of the limb-darkening theoretic models. In Sect.~\ref{sec_ttv} we
present the TTV data derived for our $17$ targets and results of their analysis, including
a detailed discussion of possible TTV trends in WASP-12 and WASP-4. In
Sect.~\ref{sec_rvttv} we present results of self-cosistent fits using both the transit and
radial velocity data, available for $13$ targets. In Sect.~\ref{sec_wasp12} we discuss in
yet more detail the case of WASP-12, deriving a purely tidal part in its observed TTV
trend.

\section{Photometric and Doppler data}
\label{sec_src}
The EXPANSION project performs a long-term monitoring of exoplanetary transits. Amateur and
professional observatories from Russia, Europe, North and South Americas with relatively
small telescopes from 25~cm to 2~m are used in the photometric observations
\citep{Sokov18}. We used data from this network, including all the data from ETD that were
used in \citep{Baluev15a}. Additionally, we used lightcurves published in the literature or
kindly provided by the observers, as listed in Table~\ref{tab_ref}. Most of them are
available in the VIZIER database.

We expanded our targets list by seven exoplanets: Qatar-2, WASP-3, -6, -12, HAT-P-3, -13,
and XO-5, thus increasing their number to $17$. The total amount of the input data has
grown considerably. This time we had $\sim 3\times 10^5$ photometric measurements in $\sim
1000$ lightcurves, compared to $\sim 8\times 10^4$ measurements in $\sim 300$ lightcurves
processed by \citet{Baluev15a}.

Whenever necessary, the timestamps in the photometric series were transformed to the ${\rm
BJD}_{\rm TDB}$ system by means of the public IDL software developed by \citet{Eastman10}.
To perform this reduction, we used ICRS coordinates through the SIMBAD database which
originate from GAIA DR2 \citep{Gaia18}. We did not apply any correction to these
coordinates due to proper motion, since this would imply only a negligible correction to
the time (below $\sim 0.1$~sec).

\begin{table}
\tiny
\caption{Sources of the photometric data (not including the EXPANSION project).}
\label{tab_ref}
\begin{tabular}{llp{31mm}}
\hline
Target    & References              & Note \\
\hline
CoRoT-2   & \citet{Gillon10}\\
\hline
GJ~436    & \citet{Gillon07}\\
          & \citet{Bean08b}         & HST Fine Guidance Sensor\\
          & \citet{Shporer09}\\
          & \citet{Caceres09}       & Very high cadence; we binned these data to $10$~sec chunks\\
\hline
HAT-P-3   & \citet{Torres07}\\
          & \citet{Chan11}\\
          & \citet{Nascimbeni11a}   & Data initially uploaded to VIZIER were not actually
                                      in BJD system as claimed (priv.~comm.); correct data uploaded in 2017 \\
          & \citet{Mancini18}\\
\hline
HAT-P-13  & \citet{Bakos09}\\
          & \citet{Szabo10}\\
          & \citet{Nascimbeni11b}\\
          & \citet{Fulton11}\\
          & \citet{Southworth12}\\
\hline
HD~189733 & \citet{Bakos06}\\
          & \citet{Winn07a}         & T10APT data involve \emph{double} HJD correction by mistake (priv.~comm.)\\
          & \citet{Pont07}          & HST Advanced Camera for Surveys\\
          & \citet{McCullough14}    & HST Wide Field Camera 3\\
          & \citet{Kasper19}        & Multi-band transmission spectroscopy; very high accuracy data\\
\hline
Kelt-1    & \citet{Siverd12}\\
          & \citet{Maciejewski18a}\\
\hline
Qatar-2   & \citet{Bryan12}         & It is not fully clear, whether the ``BJD'' times are given in UTC or TDB system.
                                      We assume BJD TDB, because the TTV residuals look bad otherwise.\\
          & \citet{Mancini14}\\
\hline
TrES-1    & \citet{Winn07b}\\
\hline
WASP-2    & \citet{Southworth10}    & Danish telescope timings might be unreliable \citep{Nikolov12,Petrucci13}\\
\hline
WASP-3    & \citet{Tripathi10}\\
          & \citet{Nascimbeni13}\\
\hline
WASP-4    & \citet{Wilson08}\\
          & \citet{Gillon09a}\\
          & \citet{Winn09}          & Superseded by \citet{Sanchis-Ojeda11}\\
          & \citet{Southworth09b}   & Danish telescope timings might be unreliable \citep{Nikolov12,Petrucci13}\\
          & \citet{Sanchis-Ojeda11}\\
          & \citet{Nikolov12}\\
          & \citet{Petrucci13}      & These data were kindly provided by the authors\\
\hline
WASP-5    & \citet{Southworth09a}   & Danish telescope timings might be unreliable \citep{Nikolov12,Petrucci13}\\
\hline
WASP-6    & \citet{Gillon09b} \\
          & \citet{TregloanReed15}\\
\hline
WASP-12   & \citet{Hebb09}          & These data were kindly provided by the authors\\
          & \citet{Chan11}\\
          & \citet{Maciejewski13}   & Partly superseded by \citet{Maciejewski16}\\
          & \citet{Stevenson14}     & Multi-band transmission spectroscopy; very high accuracy data\\
          & \citet{Maciejewski16}\\
          & \citet{Maciejewski18a}\\
\hline
WASP-52   & \citet{Chen17}          & Multi-band transmission spectroscopy; very high accuracy data\\
          & \citet{Mancini17}\\
\hline
XO-2N     & \citet{Fernandez09}\\
          & \citet{Kundurthy13}\\
          & \citet{Damasso15}\\
\hline
XO-5      & None & \\
\hline
\end{tabular}
\end{table}

Additionally, we used the precision radial velocity (RV) measurements obtained from the
archival spectra of the HARPS, HARPS-N, SOPHIE, and HIRES spectrographs. This involves the
following targets from our photometry sample: Corot-2, GJ~436, TrES-1, WASP-2, -4, -5, -6,
-12, HD~189733, XO-2N. The spectra were processed with the HARPS--TERRA pipeline
\citep{AngladaEscudeButler12}. Some of these data represent reprocessed versions of the RV
data available in the literature, e.g. from \citep{Baluev15a}, and some are new. Whenever
performing a self-consistent transit and radial velocity analysis we transform all the
Doppler time stamps to the ${\rm BJD}_{\rm TDB}$ system consistent with the photometry.
However, the RV data that we release here correspond to the UTC rather than TDB system (as
traditionally adopted for this type of the data).

Since \citet{Wilson08}, additional $31$ RV measurements have been obtained for WASP-4 with
the high resolution spectrograph CORALIE on the Swiss $1.2$~m Euler telescope at La Silla
Observatory, Chilie \citep{Queloz01}. RVs were re-computed for the new data and the dataset
presented in \citet{Wilson08}, for $45$ measurements in total, by cross-correlating each
spectrum with a G2 binary mask, using the standard CORALIE data-reduction pipeline.

For WASP-2, WASP-3, WASP-4, WASP-12, and XO-5 additional HIRES observations were presented
by \citet{Knutson14}, which we included in the analysis in the published form. The Keck RV
data from \citep{Knutson14} for XO-2N and GJ~436, and from \citep{Albrecht12} for GJ~436
were not used as they were found in our TERRA-processed sample. The HAT-P-13 data available
in \citep{Knutson14} mysteriously appeared older and much less complete than RV data set by
\citet{Winn10}, so we used the latter one. Some more in-transit RV data for WASP-12 are
also mentioned in \citep{Albrecht12} but not published.

The data files containing the photometric and radial-velocity measurements are attached as
the online-only material. The format of the files follows that of \citep{Baluev15a}.
Concerning the RVs, we currently release only a partial set, since we still plan to seek
more RV data and perform their more detailed analysis in a future work.

We notice that some TERRA-processed RV data in \citep{Baluev15a} appeared partly erratic.
First, the HARPSN data for HD~189733 appeared entirely wrong because they belong to its
known companion B. Secondly, the difference between the new and old HARPS data for GJ~436
revealed a clear systematic trend indicating some processing error in the old data set. The
long-term trend was highly significant in the previous RV release, but now it disappeared.

\section{Deriving transit times from photometry}
\label{sec_ttvmethod}
Our derivation of transit timing variations from photometry uses a similar procedure to
that of \citet{Baluev15a} which we updated to follow the processing stages below.
\begin{enumerate}
\item Fit the raw transit photometry and the resulting resulting transit timings with a
reference TTV model (linear ephemeris plus a possible quadratic trend, see
eq.~(\ref{trend})).
\item Clean TTV outliers (bad lightcurves) by verifying the TTV residuals and then
reprocessing the remaining data.
\item Clean photometry outliers in the remaining lightcurves in a similar way and then
reprocess the data.
\item Reprocess the data using semi-empirical limb-darkening coefficients for lightcurves
in which the limb-darkening was ill-fitted or had poor accuracy. The semi-empirical values
are based on \citet{ClaretBloemen11} corrected for the systematic biases derived in
Sect.~\ref{sec_ld}.
\item Among the remaining lightcurves, identify higher-quality (HQ) ones, and reprocess
them separately.
\end{enumerate}
We note that in our previous work we were only able to follow Stages~1 and~3. Stage~2 could
not be completed due to a relative lack of TTV data. Stage~4 was not performed due to a
simplistic limb-darkening treatment, which is now revised, and Stage~5 was absent. Most of
the analysis was performed using the {\sc PlanetPack} software \citep{Baluev13c,Baluev18c}.
We now consider each stage in more detail.

\subsection{Stage 1: lightcurve fitting}
The light curve fitting is based on maximum-likelihood fitting with a dedicated model of
the photometric noise and follows \citep{Baluev15a}. As in that work, we use circular model
of the curved transiter orbital motion. Most of our targets do not have a detectable
orbital eccentricity, except for GJ436b. However, the photometric data for GJ436 appeared
mostly of a too low quality. Except for a very few space-based HST observations, they do
not justify the use of a general Keplerian model. In any case, we include non-zero orbital
eccentricities in the joint transit+Doppler analysis below.\footnote{The WASP-6b nonzero
eccentricity $e\sim 0.05$, reported by \citet{Gillon09b}, is not confirmed by our joint
fits below.}

The initial steps of the algorithm involve a set of preliminary fits, needed to avoid
pathological solutions and fitting traps:
\begin{enumerate}
\item Fit the data with a fixed transit impact parameter, fixed limb darkening
coefficients and with a strictly quadratic TTV ephemeris. Contrary to \citep{Baluev15a},
who adopted a linear TTV ephemeris, here we decided to use a quadratic one because now we
have at least two candidates with a quadratic TTV trend (WASP-12 and WASP-4), and all other
targets should be processed homogeneously.
\item Refit after releasing the transit impact parameter and mid-times.
\item Refit after releasing limb darkening coefficients (except for those that are fixed at
the corrected theoretical values at Stage~4 or~5).
\item Determine very high-quality lightcurves that allow independent fitting of the limb
darkening coefficients and if such lightcurves exist, refit the model yet again.
\end{enumerate}

After these initial stages, our red noise auto-detection sequence follows that of
\citealt{Baluev15a,Baluev18c}. Our criteria for a robust red noise detection were: (i) the
log-likelihood ratio statistic $Z$ should be at least $2$, implying the asymptotic false
detection probability $\chi^2_2(Z) = \exp(-Z)\sim 14$ per cent, (ii) the uncertainty in the
red jitter $\sigma_{\rm r}$ is at most the estimated value (iii) the uncertainty in the red
noise timescale $\tau$ is at most twice the estimated value. These criteria appear very
mild (even more mild than in \citealt{Baluev15a}). In fact, they assume that most of the
lightcurves \emph{must} contain some red noise by default, except for the cases whenever
the red noise could not be modelled reliably.

In this work we used $3$ starting initial values for $\tau$, thus running up to $3$ probe
red-noise fits for each lightcurve. These initial values were spreaded logarithmically in
the range from $T/N$ to $T$ (where $T$ is the total time span of the lightcurve, and $N$ is
the number of its photometric measurements). In \citep{Baluev15a} just a single initial
value $\tau=T/\sqrt N$ was used, with a single probe fit per a lightcurve. It appeared that
among our $\sim 1000$ lightcurves, almost all reveal their red noise after just this very
first trial fit. However, in a few cases it appeared that the first fit did not converge to
a robust solution because the actual best fitting value of $\tau$ was too far from $T/\sqrt
N$. By adding two more probe fits starting from $\tau$ closer to the low and upper limits
of the range, we could robustly detect the red noise in several lightcurves additionally.

But even with these very mild detection criteria and multiple trial fits it appeared that
only $1/4$ to $1/3$ of our lightcurves (depending on the target) revealed an individually
fittable red noise. This is in agreement with \citet{Baluev15a}, however such a low
fraction of the red-noised lightcurves still appears surprising. The red noise may exist in
the rest of lightcurves too, but with ill-fitted individual parameters. Therefore, leaving
the noise models in such a partial model-mixed state might make the resulting TTV data less
homogeneous. For example, the uncertainties in the white-noise portion of TTV data may
appear systematically smaller than in the red-noise one. To soften this effect we tried to
fit the red noise in the remaining lightcurves in an averaged sense. Since the most
uncertain and poorly determinable red noise parameter is $\tau$, we assumed that this
$\tau$ is the same among all the lightcurves that did not reveal an individually detectable
red noise. While binding $\tau$ at such a shared `average' value, the value of $\sigma_{\rm
r}$ was still assumed individually fittable for each lightcurve to allow an adaptive match
of the red noise magnitude. In this way, if this derived shared $\tau$ appeared
inconsistent with the actual observations in a given lightcurve then this $\tau$ could be
just ignored by reducing $\sigma_{\rm r}$ to zero.

After that the fraction of lightcurves enclosed by a red-noise model was raised to $50-80$
per cent, depending on the target. The rest of the data had the best fitting $\sigma_{\rm
r}=0$, implying that they contradict either the derived shared $\tau$, or the red-noise
hypothesis itself. This might formally suggest the presence of a \emph{blue} noise instead
(or $\sigma_{\rm r}^2<0$). If the red noise infers an increase of the TTV uncertainties,
the blue noise would reduce them below the level expected from the white noise. Such an
apparent effect may appear due to starspot transit events (see below), but they might also
imply large individual timing biases which we do not detect or reduce in this work. In such
circumstances, we do not allow the TTV uncertainties to decrease below their white-noise
estimations.

\begin{figure*}
\includegraphics[width=0.33\linewidth]{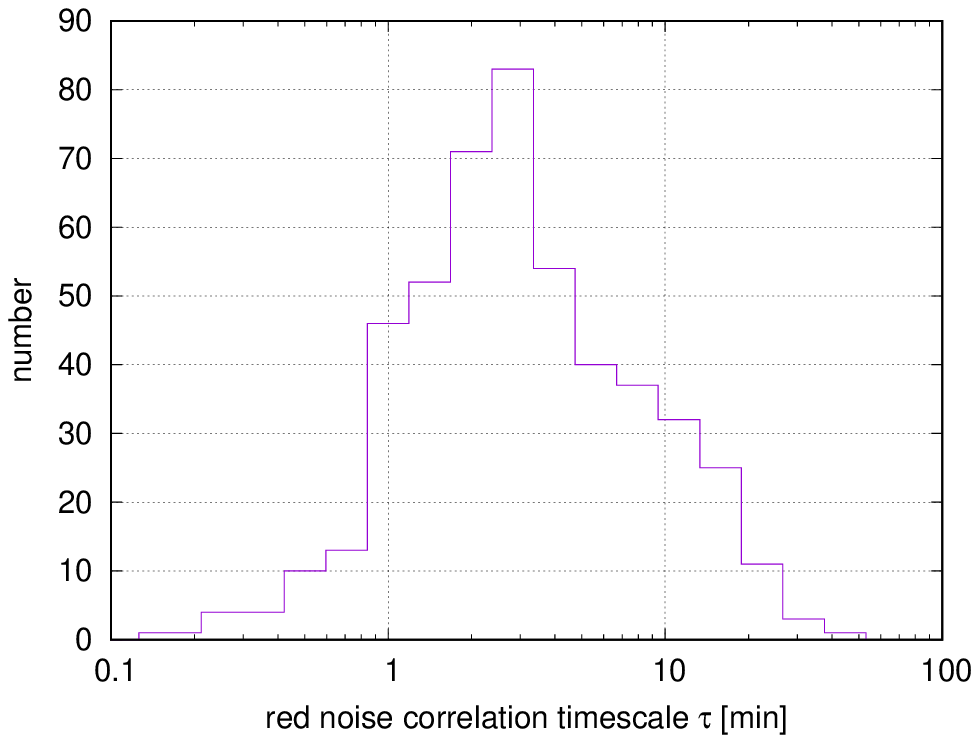}
\includegraphics[width=0.33\linewidth]{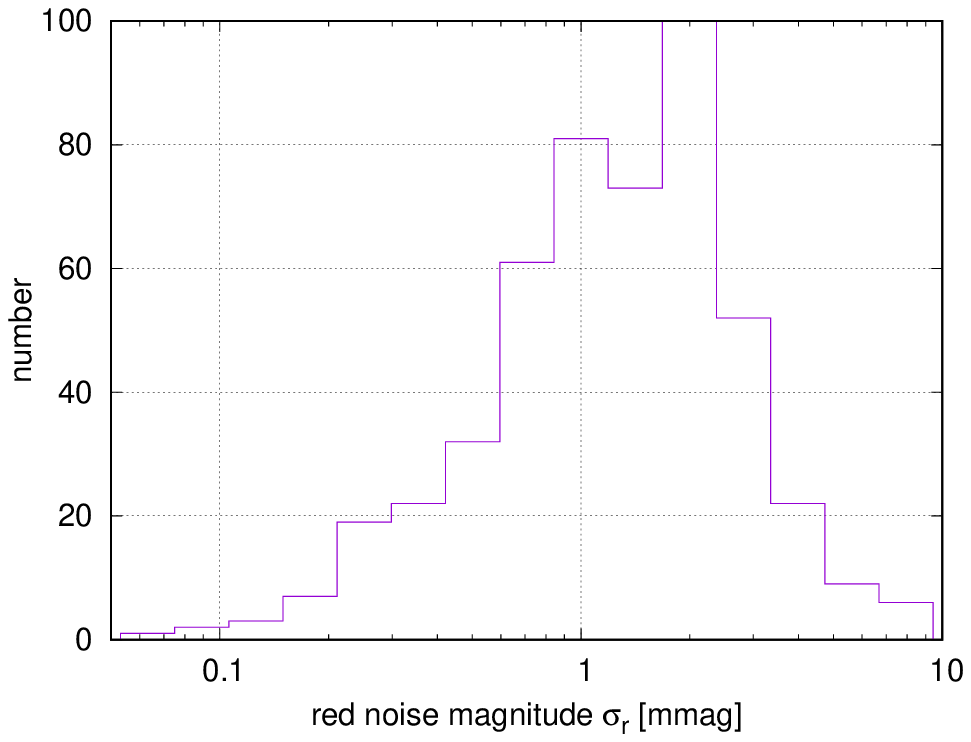}
\includegraphics[width=0.33\linewidth]{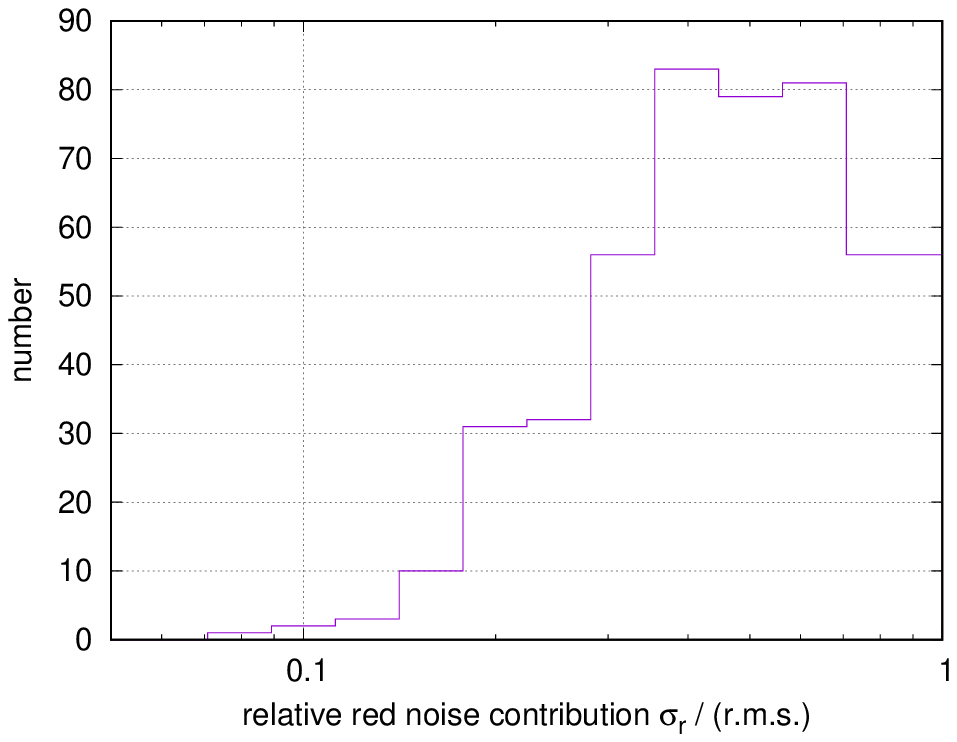}
\caption{Histograms of the estimated photometric red noise parameters,
$\tau$ and $\sigma_{\rm r}$, and of the relative red noise contribution in the total RMS.
We used only those estimations of $\tau$ and $\sigma_{\rm r}$ on Stage~4 that exceeded
their respective uncertainty.}
\label{fig_rn}
\end{figure*}

Since we have a large set of red noise estimations for numerous lighcurves, it is now
possible to consider some statistics. In Fig.~\ref{fig_rn} we show the histograms of the
derived red noise parameters $\tau$ and $\sigma_{\rm r}$, and of the ratio $\sigma_{\rm
r}/$(RMS), the relative red noise contribution in the total error budget. We can see that
$\tau$ spans a wide range from $\sim 10$~sec to $\sim 50$~min, but is primarily located in
the range $1-5$~min. The typical magnitude of the red noise is $\sim 1$~mmag, but also can
deviate a lot from this peak value. The relative red noise contribution is typically above
$30$ per cent (smaller values typically cannot be detected or estimated reliably, so they
are mostly ignored in these histograms).

Yet another major difference from \citep{Baluev15a} is a more careful treatment of the limb
darkening. As before, we adopted a quadratic limb darkening model:
\begin{equation}
I(\rho)/I(0) = 1 - A (1-\mu) - B (1-\mu)^2, \quad \mu=\sqrt{1-\rho^2},
\label{bright}
\end{equation}
where $\rho$ is the projected distance from the disk center, and coefficients $A$ and $B$
should satisfy the constraints
\begin{equation}
A+B\leq 1, \quad A+2B\geq 0, \quad A \geq 0,
\label{ldcon}
\end{equation}
which guarantee that $I(\rho)$ never turns negative and always remains monotonically
decreasing (no limb brightening allowed), see \citealt{Baluev15a,Kipping13}.

In \citep{Baluev15a} the limb darkening coefficients $A$ and $B$ were assumed the same for
the most of the lightcurves, regardless of the spectral band. But now we considered this as
an inadmissibly rough assumption. Although a fully independent fit of these coefficients
for every lightcurve is unnecessary (and even practically impossible), we need at least to
fit them independently for different spectral filters.

We split all available lightcurves into several sets that correspond to the same or similar
spectral bands. For example, we combined in a single set the Johnson $R_J$ and Cousins
$R_C$ filters, as well as the Sloan $r$ or $r'$ ones, treating them all as the same
``generic $R$'' filter. The theoretically predicted limb darkening coefficients appear
almost equal in all these filters: the differences are smaller than e.g. those implied by
different models of stellar atmosphere in \citep{Claret00,Claret04,ClaretBloemen11}. Thus
we sorted all our data into 8 classes, corresponding to the following ``generic'' spectral
ranges: $U$, $B$, $V$, $G$, $R$, $I$, $Z$, $K$. Many lightcurves (mostly amateur ones) were
obtained without any filter at all, or using a wide-band IR-UV cut-off filter, and we
joined all such data under another class labelled ``clear''.

A few lightcurves could not be assigned to any of the above band classes, because they were
obtained in another spectral band or with a different technique. Most of that data appeared
of an exceptional quality, so we always fit their limb darkening coefficients
independently. These special cases include observations from Hubble Space Telescope,
``white'' lightcurves from transmission spectroscopy, and data from some other specialized
instruments.

Finally, we carried out a systematic comparison of the resulting ``observed'' coefficients
$A$ and $B$ with their theoretically predicted values based on
\citep{Claret00,Claret04,ClaretBloemen11}. This comparison revealed certain systematic
biases, discussed in Section~\ref{sec_ld} below.

\subsection{Stages 2 and 3: cleaning the outliers}
\label{sec_st23}
The cleaning of outliers is performed as in \citep{Baluev15a}, by means of inspecting the
Gaussian quantile-quantile (QQ) plots of the TTV residuals. The QQ plot is a non-linearly
re-scaled graph of the empirical cumulative distribution $F_{\rm emp}(\varepsilon)$ of the
normalized residuals $\varepsilon_i=r_i/\sigma_i$, where $r_i$ is the best-fit residual and
$\sigma_i$ is the modelled standard deviation.\footnote{Here we assumed the multiplicative
noise model (see below) without red noise.} If the input data were good and all models
correct, then this $F_{\rm emp}(\varepsilon)$ should be close to standard Gaussian,
$F(\varepsilon) = \Phi(\varepsilon)$. Hence the quantile function $Q(\varepsilon) =
\Phi^{-1}(F_{\rm emp}(\varepsilon))$ should be close to $Q(\varepsilon)=\varepsilon$.

The graph of $Q(\varepsilon)$ is the QQ plot that we examine. These plots are given in the
online-only Fig.~2, 1st row. We can see that the empirical curves are indeed close to the
main diagonal, suggesting mostly Gaussian noise, but a number of outliers deviate in the
tails much more than a normal distribution would allow. Therefore, the outliers can be
identified as points that reside in these tails. The photometric outliers are detected in
the same way as TTV ones. The corresponding QQ plots are shown in the online-only Fig.~2,
2nd row.

We reviewed the list of potential $\sim 20$ TTV outliers, and decided to manually
`whitelist' two lightcurves looking like outliers. Namely, this is one lightcurve for
HAT-P-13 from \citep{Szabo10} and one for HD~189733 from \citep{Kasper19}, both with
$\varepsilon\simeq 4$. Concerning HAT-P-13, it demonstrated inconclusive hints of a TTV in
the past, and it might appear to be the case that \citep{Szabo10} measurements actually
reveal a true TTV, rather than a statistical outlier (e.g. induced by known non-transiting
companions, see \citealt{Winn10}). However, after that we noticed that this large
normalized residual was finally reduced on Stage~4, thanks to using corrected
limb-darkening coefficients which appeared ill-fit on Stage~3. Concerning the HD~189733
lightcurve by \citet{Kasper19}, it belongs to a homogeneous set of high-quality
transmission spectroscopy observations. The other observations also have rather large
$\epsilon$ level. We decided to allow all the \citet{Kasper19} data to Stage~5 despite the
particular lightcurve being rather anomalous. Possible reasons of such anomalies in the
\citet{Kasper19} data are discussed in Sect.~\ref{sec_qttv}.

\subsection{Stage 4: applying empirically corrected limb-darkening}
Many lightcurves have relatively poor quality, so it is not possible to reliably fit a
two-parametric limb-darkening law~(\ref{bright}). Therefore, on Stage~3 multiple estimates
appear to have large uncertainties in $A$ and $B$ about unity, or the coefficients
themselves lie on the boundary of their admissible domain \citep{Kipping13}, indicating a
poor fit. To overcome these issues, we performed one more processing pass, fixing the
limb-darkening coefficients with poor accuracy at certain semi-empirical values. See more
detailed discussion and motivation in Sect.~\ref{sec_ld}.

\subsection{Stage 5: determining high quality lightcurves}
To identify transit times of a higher quality, we first introduce the ``quality
characteristic'' of a lightcurve:
\begin{equation}
\mathcal Q = \frac{\sqrt{\text{measurements density}}}{\text{residuals RMS}}
\end{equation}
The quantity $1/\mathcal Q$ determines the uncertainty offered by a ``standard'' chunk of
the lightcurve of a unit length. The uncertainty of an arbitrary chunk of length $t$ scales
as $1/(\mathcal Q\sqrt t)$. Here we neglect possible red noise, so even neighboring
measurements are assumed uncorrelated.

This characteristic is not yet indicative concerning a particular exoplanet. Let $\tau$ be
the transit duration, and $r'=r_{\rm pl}/R_\star$ be the planet/star radii ratio. Then the
uncertainty of the in-transit piece of the lightcurve would be $1/(\mathcal Q
\sqrt{\tau})$, and it should be compared to the transit depth $r'^2$. That is, the
following normalized parameter:
\begin{equation}
\mathcal Q' = \mathcal Q \sqrt{\tau} / r'^2
\end{equation}
can serve as our idealized quality characteristic. Say, $\mathcal Q'=100$, then the transit
depth can be measured with an accuracy of $1$ per cent, while $\mathcal Q'=10$ implies
relative accuracy of $10$ per cent.

Now let us plot the empirical distribution of $\mathcal Q'$ computed for all our lightcurves
(online-only Fig.~2, bottom row). We can see that $\mathcal Q'$ varies in an very wide
range from a few tens to a few thousands. We choose a threshold $Q'>100$ to select the HQ
lightcurves. Such a threshold keeps about $2/3$ of the entire sample, so it is a relatively
mild filter. Our goal was mainly to filter out only very inaccurate and probably useless
data, rather than to select a minor portion of highly accurate ones.

Note that whenever a lightcurve has low $\mathcal Q'$, this does not necessarily mean that
it must be immediately removed from the analysis as unreliable. Such a lightcurve just has
a poor overall accuracy, but it already survived the normality tests of the previous
processing stages. Statistically, the derived timing value remains quite admissible and
usable (within its uncertainty). Below we consider results of Stage~4 and Stage~5
simultaneously so the reader can compare them.

\section{Empirical calibration of the limb-darkening coefficients}
\label{sec_ld}
Using the technique presented above, we performed a per-target and per-band fit of the limb
darkening coefficients $A$ and $B$ in the quadratic model~(\ref{bright}). After that, we
compared these empirical estimates with their theoretically predicted values from
\citep{Claret00,Claret04} and their update from \citep{ClaretBloemen11} band-by-band. The
coefficients for the ``clear'' band-class were compared with the bolometric estimates by
Claret. We utilized the {\sc jktld} code by \citet{Southworth15} that offers a convenient
interface for interpolating the original tables by Claret.\footnote{See
\url{http://www.astro.keele.ac.uk/jkt/codes/jktld.html} for download; we actually augmented
this code to process the newer tables by \citet{ClaretBloemen11}, and applied an additional
post-interpolation with respect to the metallicity, which is merely selected rather than
interpolated by {\sc jktld}.} The necessary stellar parameters ($T_{\rm eff}$, $\log g$,
[Fe/H]) were taken mainly from the SWEET-Cat \citep{Santos13}, and from \citep{Siverd12}
for Kelt-1. In almost all cases we used the coefficients corresponding to the ATLAS models,
except for GJ~436, for which only the PHOENIX-based coefficients were available. We assumed
microturbulence velocity $v_t$ of $2$~km/s for all cases.

We found that many of our $A$ and $B$ estimations, even the most accurate ones,
significantly deviate from theoretical values. In itself, this is not very surprising,
because the theoretical values are expected to have some biases \citep{Heyrovsky07}. Even
if the theoretical brightness profile was entirely perfect, the two-parameter models such
as~(\ref{bright}) cannot approximate it everywhere equally well. The resulting
``theoretical'' coefficients $A$ and $B$ depend on how we fit this profile: they may appear
biased to better fit one its portion or another. And they should not necessarily coincide
with the empirical values obtained from transit fitting (even if the latter had no
significant errors at all).

In online-only Fig.~3, some worst-case discrepancies are demonstrated. The empirical $A$
and $B$ estimations correspond to the processing Stage~3, while the theoretical values were
derived from \citep{Claret00,Claret04}, and one can see that they systematically deviate by
$\sim 0.1-0.2$.

Then we computed the universal shifts $\Delta A$ and $\Delta B$, necessary to minimize the
differences between the observed and theoretical coefficients. The weighted least squares
fit yielded the biases $\Delta A=0.059\pm 0.008$ and $\Delta B=-0.172 \pm 0.014$ for the
quadratic law, and $\Delta A=-0.112\pm 0.002$ for the linear law. These shifts refer to the
older tables by \citet{Claret00,Claret04}, ATLAS models, and take into account only the
UBVGRIZK filters.

By fitting the newer models by \citet{ClaretBloemen11}, corresponding to the
flux-conservation method (FCM), and for the same spectral filters as above, we obtained the
following biases: $\Delta A=0.004\pm 0.008$ and $\Delta B=-0.099 \pm 0.014$ for the
quadratic law, and $\Delta A=-0.035\pm 0.002$ for the linear law. The newer tables are
clearly better, though some minor bias still remains in $B$. By adding the latter
best-fitting corrections to the theoretical $A$ and $B$ values the agreement can be
improved remarkably. This becomes obvious in several high-accuracy cases (e.g. WASP-4,
Qatar-2), see online-only Fig.~4.

The coefficients from \citet{ClaretBloemen11} obtained by least-square fit of the
brightness profile appear less accurate than the FCM ones and more similar to those from
\citep{Claret00,Claret04}. The differences between various systems of the limb-darkening
coefficients highlight the need for a homogeneous TTV analysis, based on simultaneous
fitting of all raw lightcurves at once and using the same analysis pipeline. Direct mixing
of independently derived timing measurements, especially those released before or after
the 2011 update, may lead to spurious timing biases.

In this work, we adopt a hybrid approach to model the limb-darkening profile following the
key aspects below.
\begin{enumerate}
\item If at Stage~3 both $A$ and $B$ had a fitting uncertainty of better than $0.2$ and
simultaneously did not reside on either boundary of~(\ref{ldcon}) then we did not rely on
the theoretical values. Even the corrected ones may still appear to be biased for an
individual star, so we allowed these coefficients to be fitted from the transit curves as
free parameters (still taking into account the common binding constraints per each spectral
band class).

\item The limb-darkening coefficients corresponding to filters other than UBVGRIZK, were
always fitted, including the no-filter (``Clear'') cases, regardless of their resulting
accuracy. Notice that in the online-only Figs~3 and~4 we compare the ``Clear'' band with
the predicted bolometric values only for a reference: we do not rely on the bolometric
coefficients in our processing.

\item If at Stage~3 either limb-darkening estimations appeared too uncertain
(above $0.2$) or the model appeared ill-fitted (residing at the boundary of~(\ref{ldcon}))
then we fixed such coefficients $A,B$ at their theoretical FCM values from
\citep{ClaretBloemen11}, corrected by the biases derived above. This refers to only the
UBVGRIZK filters. The motivation here was to get rid of unrealistic solutions.
\end{enumerate}

The graphs of the final limb-darkening coefficients are presented in the online-only Fig.~5.

\section{Results of the transits analysis}
\label{sec_ttv}
\subsection{Verifying the quality of the derived timings}
\label{sec_qttv}
Before presenting our TTV analysis results, we need to discuss the quality of the derived
transit timing data. Our transit analysis pipeline differs in several important aspects
from the standard methods applied usually. In particular, we treat the red noise using a
parametric model by Gaussian processes with exponential correlation function. While many
other works may use different techniques, e.g. originating from a seminal work by
\citep{Pont06} or from \citep{Foreman-Mackey17}. Also, we used different statistical
treatment paradigms and different software. Finally, we analyse jointly lightcurves of a
very different quality, from amateur ones to professional ground-based and even space-based
HST data. Although we undertook multiple efforts to handle such a heterogeneity, its side
effects may still exist.

Therefore, we need some benchmark of the accuracy and quality of our TTV data. This can
be done by comparing them with analogous TTV data from other published works. However, most
of the published TTV data were derived by different teams who used different techniques and
different assumptions (e.g. concerning the limb-darkening). Hence, their mixed compilations
cannot usually serve as reliable benchmarks. We need a long record of TTV data, obtained
mostly by the same team.

In our target list, only WASP-12 perfectly suits our needs. It has $\sim 200$ observed
transit lightcurves in total, and about half of them were processed by the same group
\citep{Maciejewski13,Maciejewski16,Maciejewski18a}. Simultaneously, these data were obtained
at quite different telescopes located in different astroclimate conditions. Therefore, they
have different quality characteristics, offering the necessary degree of physical
heterogeneity.

Here we used the transit times from \citep{Maciejewski18a} and \citep{Maciejewski16} that
included the most reprocessed lightcurves of \citep{Maciejewski13}. We did not include the
timings from the 2013 paper not reprocessed in the 2016 one (since they would be
statistically different). After that we sampled the same transits from our homogeneous data
release. Thus we obtained two similar TTV time series to be compared, each containing $73$
data points at the same epochs. We fitted both them with a quadratic trend
model~(\ref{trend}), resulting in almost identical trend fits. We then computed the
resulting RMS: $22.1$~sec for the Maciejewski~et~al. data and $24.9$~sec for our data
release. The Maciejewski~et~al. data winning with a slightly smaller scatter of the
residuals, though this $12$ per cent difference is comparable to the probable statistical
uncertainty ($1/\sqrt N$ for $N=73$ yields the same $12$ per cent). Therefore, the
intrinsic statistical accuracy of our processing pipeline appears similar high quality TTV
data sets available in the literature.

However, our data reveal important difference in another aspect. The value of the reduced
$\chi^2$ for the quadratic TTV model is $0.79$ for Maciejewski~et~al., implying that they
tend to overestimate their TTV uncertainties by $12$ per cent on average. On the contrary,
our data imply the reduced $\chi^2$ of $1.49$, which means that our TTV uncertainties
appear underestimated by the factor $\sqrt{1.49}$, or by $\sim 22$ per cent on average.

We notice that it is quite frequent that the uncertainties reported for some measurements
have a remarkable systematic bias. This is expected, because there are always subtle
physical effects that were missed, or shortcomings of the adopted models, or hidden
inaccuracies of the statistical processing. All this may lead to a systematically wrong
uncertainties in the derived data. The same phenomenon was known long ago in the precision
Doppler data \citep{Wright05,Baluev08b}, and a generally similar effect should be expected
in TTVs.

Then the TTV data released by different groups may have quite a different level of hidden
noise. We therefore caution the reader against simplistic joining of TTV data coming from
different sources. Such a merging should be made in an adaptive manner instead, taking into
account possibly different relative weights of heterogeneous subsets. One way of such an
adaptive treatment is demonstrated below for the WASP-4 case.

As we can see, the TTV noise uncertainties may appear overestimated (like in
Maciejewski~et~al.), as well as underestimated (like in this work). Concerning the first
case, the data have a smaller actual scatter than expected, indicating just some
unclassified inaccuracies in the processing algorithm. Concerning the second case, this can
be also explained in a bit more physical manner via the effect of an additional noise
source, not taken into account when performing the processing.

We believe that this source can be the starspot transit events. Initially, we expected that
such transit curve anomalies might be taken into account by a red noise model, however it
appeared that lightcurves with obvious spot-transit anomalies usually do not have a
detectable or even fittable red noise. Moreover, in practice it sometimes appeared that
such transit curves demonstrated hints of a \emph{blue} noise with $\sigma_{\mathrm
r}^2<0$.

One may argue that such a behaviour is reasonable. The type of the noise~--- white, red, or
blue~--- is basically determined by its rate of decrease whenever it is averaged over $N$
consequent observations: either $N^{-\frac{1}{2}}$ (white), or slower than that (red), or
quicker than that (blue). A single spot-transit perturbation in the lightcurve is actually
not noise: it is a determenistic curve anomaly. The noise-like effect here appears only
because these anomalies change randomly from one transit to another. However, for a given
lightcurve any spot-transit anomaly behaves as a deterministic function, e.g. it is
averaged out at the rate $1/T$, where $T$ is the length of the observation sequence. This
corresponds to the decay rate of $1/N$, if $N$ is accumulated linearly with time.
Therefore, such an anomaly can be interpreted as a blue noise rather than red or white one.

In particular, we notice that some HQ observations by \citet{Kasper19} may be affected by
hidden starspot transit or other activity-related phenomena (even though they are not
obvious from the lightcurve, possibly due to a low cadence). This might explain why one of
them was identified as an outlier deviating by $\sim 2$~min (see Sect.~\ref{sec_st23}).
Note that our estimation of this transit time is essentially consistent with the original
\citet{Kasper19} value (the shift by just $6$~sec, our uncertainty is $12$~sec compared to
the original uncertainty of $11$~sec), so this issue cannot be attributed to our
data-analysis pipeline. The star HD~189733 itself reveals a remarkably large scatter of the
TTV residuals (see Table~\ref{tab_traw} explained below), possibly indicating an increased
starspot activity.

\subsection{Analysis of the TTV}
We processed our TTV data in the homogeneous manner, using the same protocol for each
target. For the first step, we tested the existence of a possible long-term nonlinear trend
in the TTV time series, expressing it as a quadratic model:
\begin{equation}
\text{TTV}(n) = T_0 + P (n-n_0) + \frac{1}{2} \frac{dP}{dn} (n-n_0)^2,
\label{trend}
\end{equation}
where $n$ is the transit count (or epoch), $P$ is the orbital period, and $dP/dn$ is the
small quadratic coefficient. Defining a temporal variable $t=nP$, we can alternatively
rewrite~(\ref{trend}) as:
\begin{eqnarray}
\text{TTV}(t) &=& T_0 + (t-t_0) + \frac{1}{2} \frac{\dot P}{P} (t-t_0)^2 \nonumber\\
              &=& T_0 + (t-t_0) - \frac{(t-t_0)^2}{2T_{\rm d}}.
\end{eqnarray}
In this model the quantity $T_{\rm d}=-P/\dot P$ represents a characteristic time of period
decay (the time when the apparent period would turn zero if it decreased linearly). Since
it has an intuitive interpretation, we often use this quantity below as a reference fit
parameter (rather than the quadratic coefficient itself). However, we emphasize that
multiple physical phenomena may be approximated by mathematically the same
formula~(\ref{trend}): tidal orbital decay, tidal apsidal drift, or even non-tidal effect
of a perturbation from a distant companion (causing the TTV via the light-travel effect).

The TTV residuals themselves are plotted in the online-only Fig.~6 and~7 for all our
targets. They correspond to a linear TTV ephemeris and are given separately for Stage~4
(all data) and Stage~5 (HQ data).

We were able to easily confirm the TTV trend of WASP-12 \citep{Maciejewski16} at this step.
This case is discussed in details below in a separate section. The TESS timing data
\citep{Bouma19} claimed that a similar TTV trend may exist in WASP-4, but our data do not
confirm such a trend. The detailed analysis of this target is discussed below in a separate
section.

The other targets did not demonstrate convincingly detectable hints of nonlinear TTV trends
(based on the log-likelihood tests applied to the TTV time series, see
Section~\ref{sec_WASP4}). Furthermore, we performed a search for periodic TTV signals. We
constructed a periodogram $z_3$ from \citep{Baluev08a}, shown in the online-only Fig.~8
(for Stage~4) and Fig.~9 (for Stage~5). The base model for this periodogram always included
a quadratic trend.

We could not find any periodic TTV for any of the targets. Peridograms did not reveal hints
of significant periodicity. In particular, we do not detect any hints of previously claimed
controversial TTV for HAT-P-13 \citep{Nascimbeni11b,Pal11} or for WASP-3
\citep{Maciejewski10,Montalto12,Maciejewski18b}. Concerning the HAT-P-13 target, it has a
second companion HAT-P-13~\emph{c}, and also reveals hints of additional long-period
companions appearing as a linear RV trend \citep{Winn10}. These additional companions would
impose a variable light travel delay effect on the inner tight subsystem, causing therefore
a TTV. However, this type of TTV is not detectable in HAT-P-13 due to the small magnitude
(e.g. $\sim 7$~sec from HAT-P-13~\emph{c}). In this work we did not investigate the TTVs
possibly coming from gravitational perturbations of the planet \emph{b} orbital motion.

For WASP-4 HQ data we find that multiple peaks rise above the two-sigma significance level
in the short-period range. However, these peaks look more like noise rather than a
systematic variation. Moreover, they disappear if we remove just a single timing
measurement, namely the one derived from the lightcurve by \citet{Sanchis-Ojeda11}, dated
by 02 Aug 2009. We believe that this lightcurve could be affected by a subtle residual
systematic effect or by a spot-transit event, even though it was not classified as an
outlier and looks visually reasonable. Similar issues may apply to HD~189733, which
involves at least one lightcurve by \citet{Kasper19} with anomalous timing.

We note that in \citep{Baluev15a} inconclusive hints of periodic TTVs for WASP-4 were
claimed in the range of a few days. However those periodogram peaks disappeared when
applying a more careful treatment of the limb-darkening coefficients. This highlights the
practical value of the limb-darkening model, even if it apparently does not seem so
important for TTV studies.

\subsection{Updated planetary transit fits}
\label{sec_traw}
The Table~\ref{tab_traw} contains fitted transit parameters for our $17$ exoplanets, both
for the Stage~4 and Stage~5 data. We give only rather raw parameters, while the complete
set can be determined only from the transit+RV fits (considered below). In addition, we
give the number of red-noise lightcurves for each target (fitted individually or with
shared $\tau$), the cumulative quality characteristic for each target, the maximum and mean
absolute correlation of the derived transit times (which appears between different transits
through the shared planetary parameters), and the reduced $\chi^2$ for the derived transit
times residuals (relative to a best fitting quadratic TTV). We also performed alternative
fits assuming that all the transit times strictly follow a quadratic model. For these
alternative fits we only consider the best fitting quadratic TTV ephemeris~(\ref{trend}).

Our approach may inspire statistical correlations between different transit times
\citep{Baluev15a}, but they mostly appeared negligible. Only for GJ~436 and WASP-6 some
pairs of transits generated a large correlation of up to $0.68$. This is because now we
included several partial transits in the analysis. Nonetheless, on average the effect of
correlations becomes negligible, so we decided to keep such transits particularly since we
have rather little transits data for these two targets.

We notice that for Kelt-1 the impact parameter estimation $b=0.05\pm 0.64$ is a formal and
non-informative value, since the parameter $b$ becomes severely nonlinear and hence
non-Gaussian whenever it becomes smaller than the uncertainty. In this case, a considerably
more linear parameter might be $a=\sqrt{1-b^2}$ with $\sigma_a = (b/a)\sigma_b$ (if $b$ is
the distance of the transit trajectory from the star disk center, $a$ is its distance from
the star limb). This corresponds to $a=0.999\pm 0.032$, implying the $1\sigma$ low limit on
$a$ of $0.967$, hence a more realistic upper $1\sigma$ limit on $b$ of $0.25$ (rather than
$0.64$). In \citet{Baluev15a} the Kelt-1 best fit would formally correspond to an $a>1$,
i.e. imaginary $b$, so it was set to the least physically sound value $b=0$. Clearly, the
value of $b$ is consistent with zero in any case, but its uncertainty still remains large.
To avoid the mathematical peculiarity near $b=0$, one could consider $a$ or e.g. $b^2$ as a
primary fit parameter, however we keep using $b$ as it is more traditional and intuitive.

Finally, the most important observation from Table~\ref{tab_traw} is that all values of
$\chi^2_{\rm TTV}$ are significantly above one. This indicates, most probably, that our
algorithm does not take into account all the noise sources in full. As we already noticed
above in Sect.~\ref{sec_qttv}, one such escaped noise source is likely the effect of
spotting activity causing random anomalies in transit lightcurves.

It is important for us that this activity effect, whatever physical source it has, can be
easily modelled at the TTV processing stage. This can be achieved by fitting an additive
noise increasing derived timing uncertainties, or by multiplying them by a constant factor
(we did not find definite hints clearly favouring either of these approaches). These
methods are discussed in detail in \citep{Baluev08b,Baluev14a}. However, all
self-consistent fits that avoid explicitly dealing with transit timings may appear to have
underestimated uncertainties because of this activity effect. This refers, in particular,
to the quadratic ephemeris given in Table~\ref{tab_traw}. For example, for WASP-12 the
relative uncertainty of $dP/dn$ following from the table is $6.7$ per cent, while after
processing the transit times with an adaptive noise model (see Section~\ref{sec_WASP12}
below) we obtain a larger relative uncertainty about $9.2$ per cent, which is more
realistic. The ratio of these uncertainties is almost equal to the value of
$\sqrt{\chi^2_{\rm TTV}}$ from Table~\ref{tab_traw}.

We expect that the values of $T_0$ and $P$ from Table~\ref{tab_traw} are affected in the
same way, as well as $P$ and $T_{\rm d}$ from Table~\ref{tab_scp} containing the
self-consistent transit+RV fits. Their uncertainties following from a self-consistent fit
should be multiplied by the factor of $\sqrt{\chi^2_{\rm TTV}}$. Concerning the other
fitted parameters, their uncertainties may also be affected, but in an unpredictable
manner. The correction factor is not necessarily related to $\chi^2_{\rm TTV}$, if the
parameter has no direct relationship with transit times.

\begin{table*}
\tiny
\caption{Fitted parameters of exoplanetary transit curves after Stages~4 and~5 processing.}
\label{tab_traw}
\begin{tabular}{@{\;}l@{\;}cc@{\;}l@{\;}l@{\;}l@{\;}l@{\;}c@{\;}c@{\;}ccl@{\;}l@{\;}l@{\;}}
\hline
             & total     & \multicolumn{8}{c}{Assuming fittable transit times}& &\multicolumn{3}{c}{Fixing timings at a quadratic model$^2$}\\
\cline{3-10}\cline{12-14}
transiter    & number of & number of & total      & radii ratio& half-duration        & impact par.          &        & \multicolumn{2}{@{}c@{}}{mid-times correl.} & &   ref. mid-time$^5$ & orbital period$^5$ & TTV trend$^5$\\
host         & transits  & red-noised&quality$^1$& $r=R_{\rm pl}/R_\star$ &$t_d/2$ [days]& $b$ & $\sqrt{\chi^2_{\rm TTV}}$ $^2$& MAD & MAX & &$T_0$ [${\rm BJD}_{\rm TDB}-$ & $P$ [days]      & $dP/dn$ \\
             & $N$       &lightcurves$^3$&$\mathcal Q'_{\rm sum}$& & &                    &        &              &              & &$-2450000$]                      &           & [$10^{-10}$day] \\
\hline
Corot-2      &  $38$ &  $13+12$ &    $893$ &        $0.16524(93)$ &        $0.04726(20)$ &          $0.158(70)$ & $1.37$ &    $0.00034$ &      $0.006$ & &     $7622.43669(11)$ &     $1.74299767(36)$ &          $2.6(5.1)$ \\
GJ436$^4$    &  $47$ &    $8+7$ &    $857$ &         $0.0847(10)$ &        $0.02108(21)$ &         $0.8612(66)$ & $1.10$ &     $0.0022$ &       $0.66$ & &     $4439.41624(10)$ &     $2.64389938(67)$ &           $-28(12)$ \\
HAT-P-13     &  $51$ &  $19+12$ &   $1470$ &        $0.08826(78)$ &        $0.06919(38)$ &         $0.7500(77)$ & $1.55$ &     $0.0011$ &      $0.036$ & &     $5476.91220(19)$ &      $2.9162394(14)$ &            $92(33)$ \\
HAT-P-3      &  $66$ &  $16+16$ &   $1400$ &        $0.11091(48)$ &        $0.04335(17)$ &          $0.615(12)$ & $1.48$ &    $0.00015$ &      $0.011$ & &     $7237.38678(10)$ &     $2.89973797(38)$ &          $8.0(9.6)$ \\
HD189733     & $106$ &  $27+32$ &  $12800$ &        $0.15703(32)$ &       $0.037514(48)$ &         $0.6646(20)$ & $2.10$ &    $0.00029$ &       $0.17$ & &    $3968.837026(20)$ &     $2.21857545(15)$ &         $-5.3(1.9)$ \\
Kelt-1       &  $34$ &  $12+14$ &    $938$ &        $0.07584(82)$ &        $0.05682(28)$ &           $0.05(64)$ & $1.91$ &    $0.00092$ &      $0.023$ & &     $8026.51487(14)$ &     $1.21749220(71)$ &        $-20.6(8.6)$ \\
Qatar-2      &  $59$ &  $12+20$ &   $1750$ &        $0.16165(91)$ &        $0.03812(11)$ &          $0.129(63)$ & $1.26$ &    $0.00021$ &      $0.014$ & &    $6045.458848(35)$ &     $1.33711643(26)$ &          $4.8(3.6)$ \\
TrES-1       &  $56$ &  $13+23$ &   $1370$ &        $0.13799(83)$ &        $0.05233(17)$ &          $0.238(45)$ & $1.17$ &    $0.00029$ &     $0.0067$ & &    $4350.354597(84)$ &     $3.03006957(34)$ &          $1.0(7.1)$ \\
WASP-2       &  $68$ &  $20+19$ &   $1530$ &        $0.13315(50)$ &        $0.03727(15)$ &         $0.7382(50)$ & $1.47$ &    $0.00024$ &     $0.0042$ & &     $5513.13577(12)$ &     $2.15222222(39)$ &          $3.0(8.0)$ \\
WASP-3       &  $69$ &  $17+20$ &   $1710$ &        $0.10637(58)$ &        $0.05700(18)$ &          $0.492(17)$ & $1.53$ &    $0.00025$ &      $0.068$ & &    $5325.825419(88)$ &     $1.84683507(26)$ &          $5.4(4.0)$ \\
WASP-4       &  $66$ &  $22+22$ &   $4250$ &        $0.15488(32)$ &       $0.044907(53)$ &          $0.130(29)$ & $1.31$ &    $0.00019$ &     $0.0094$ & &    $5045.738470(22)$ &    $1.338231531(83)$ &         $-0.98(94)$ \\
WASP-5       &  $17$ &    $9+3$ &   $1220$ &        $0.11459(79)$ &        $0.05030(22)$ &          $0.453(25)$ & $1.62$ &      $0.002$ &      $0.022$ & &     $5896.57891(15)$ &     $1.62843035(71)$ &           $-10(17)$ \\
WASP-6       &  $18$ &    $8+3$ &   $1460$ &        $0.14310(88)$ &        $0.05368(19)$ &          $0.222(50)$ & $1.68$ &      $0.006$ &       $0.68$ & &    $5379.546164(80)$ &     $3.36100260(57)$ &           $-28(22)$ \\
WASP-12      & $230$ &  $84+72$ &   $9070$ &        $0.11840(17)$ &       $0.062362(49)$ &         $0.4312(48)$ & $1.35$ &    $0.00013$ &      $0.018$ & &    $5994.401004(25)$ &    $1.091420405(51)$ &         $-9.51(64)$ \\
WASP-52      &  $72$ &  $25+20$ &   $2010$ &        $0.16538(51)$ &        $0.03875(11)$ &         $0.5985(62)$ & $1.56$ &    $0.00063$ &        $0.1$ & &    $6904.792855(60)$ &     $1.74978179(37)$ &           $-19(11)$ \\
XO-2N        &  $73$ &  $29+20$ &   $3930$ &        $0.10348(38)$ &       $0.055937(92)$ &          $0.194(40)$ & $1.44$ &     $0.0004$ &      $0.054$ & &    $5167.935634(40)$ &     $2.61585965(16)$ &         $-4.1(4.1)$ \\
XO-5         &  $28$ &    $9+4$ &    $678$ &         $0.1026(15)$ &        $0.06354(58)$ &          $0.537(40)$ & $2.12$ &    $0.00075$ &       $0.01$ & &     $7760.49334(36)$ &      $4.1877641(22)$ &           $258(67)$ \\
\hline
Corot-2      &  $25$ &   $10+9$ &    $856$ &        $0.16557(99)$ &        $0.04728(22)$ &          $0.190(63)$ & $1.55$ &    $0.00061$ &     $0.0074$ & &     $7629.40868(12)$ &     $1.74299768(47)$ &          $3.2(6.5)$ \\
GJ436$^4$    &  $11$ &    $2+0$ &    $757$ &         $0.0829(15)$ &        $0.02107(26)$ &         $0.8664(78)$ & $1.32$ &      $0.041$ &       $0.66$ & &     $4280.78227(12)$ &      $2.6439052(85)$ &         $-160(190)$ \\
HAT-P-13     &  $38$ &  $16+10$ &   $1420$ &        $0.08757(81)$ &        $0.06895(39)$ &         $0.7456(82)$ & $1.54$ &     $0.0015$ &      $0.037$ & &     $5511.90699(19)$ &      $2.9162394(14)$ &            $96(34)$ \\
HAT-P-3      &  $34$ &    $7+7$ &   $1340$ &        $0.11075(50)$ &        $0.04337(18)$ &          $0.622(12)$ & $1.54$ &    $0.00034$ &      $0.012$ & &     $7237.38674(10)$ &     $2.89973825(42)$ &            $14(11)$ \\
HD189733     &  $75$ &  $24+19$ &  $12800$ &        $0.15696(33)$ &       $0.037525(48)$ &         $0.6645(20)$ & $2.39$ &    $0.00046$ &       $0.17$ & &    $3968.837031(20)$ &     $2.21857545(16)$ &         $-5.6(2.0)$ \\
Kelt-1       &  $18$ &    $6+5$ &    $887$ &        $0.07523(89)$ &        $0.05694(31)$ &           $0.13(26)$ & $1.57$ &     $0.0018$ &      $0.019$ & &     $8026.51487(15)$ &     $1.21749273(88)$ &           $-14(11)$ \\
Qatar-2      &  $25$ &   $8+10$ &   $1720$ &        $0.16159(93)$ &        $0.03810(11)$ &          $0.125(66)$ & $1.40$ &     $0.0005$ &      $0.014$ & &    $6034.761900(36)$ &     $1.33711631(33)$ &          $4.0(5.6)$ \\
TrES-1       &  $44$ &  $14+18$ &   $1340$ &        $0.13772(85)$ &        $0.05228(17)$ &          $0.235(47)$ & $1.13$ &    $0.00036$ &     $0.0071$ & &    $4347.324507(86)$ &     $3.03006949(35)$ &          $4.2(7.4)$ \\
WASP-2       &  $32$ &  $12+10$ &   $1450$ &        $0.13328(52)$ &        $0.03730(16)$ &         $0.7378(54)$ & $1.64$ &    $0.00037$ &     $0.0059$ & &     $5405.52461(14)$ &     $2.15222181(58)$ &            $10(10)$ \\
WASP-3       &  $45$ &  $10+16$ &   $1660$ &        $0.10630(60)$ &        $0.05704(19)$ &          $0.495(18)$ & $1.52$ &    $0.00046$ &      $0.071$ & &    $5325.825409(91)$ &     $1.84683487(28)$ &          $8.2(4.2)$ \\
WASP-4       &  $50$ &  $17+19$ &   $4240$ &        $0.15488(32)$ &       $0.044915(53)$ &          $0.134(28)$ & $1.22$ &    $0.00025$ &     $0.0093$ & &    $5045.738469(23)$ &    $1.338231514(85)$ &         $-1.0(1.0)$ \\
WASP-5       &  $15$ &    $8+3$ &   $1220$ &        $0.11477(79)$ &        $0.05037(23)$ &          $0.459(25)$ & $1.71$ &     $0.0026$ &      $0.022$ & &     $5896.57894(15)$ &     $1.62843056(75)$ &           $-16(19)$ \\
WASP-6       &  $17$ &    $8+3$ &   $1460$ &        $0.14308(88)$ &        $0.05368(19)$ &          $0.220(51)$ & $1.72$ &     $0.0067$ &       $0.68$ & &    $5379.546159(80)$ &     $3.36100249(57)$ &           $-30(22)$ \\
WASP-12      & $203$ &  $78+58$ &   $9090$ &        $0.11839(17)$ &       $0.062367(49)$ &         $0.4316(48)$ & $1.33$ &    $0.00015$ &      $0.018$ & &    $6003.132366(25)$ &    $1.091420385(52)$ &         $-9.40(65)$ \\
WASP-52      &  $44$ &  $19+10$ &   $1970$ &        $0.16530(53)$ &        $0.03877(12)$ &         $0.5965(65)$ & $1.59$ &     $0.0013$ &       $0.11$ & &    $6904.792887(62)$ &     $1.74978159(40)$ &           $-14(12)$ \\
XO-2N        &  $54$ &  $24+14$ &   $3920$ &        $0.10358(38)$ &       $0.055930(92)$ &          $0.207(37)$ & $1.56$ &    $0.00069$ &      $0.054$ & &    $5212.405250(40)$ &     $2.61585963(16)$ &         $-2.5(4.3)$ \\
XO-5         &  $12$ &    $3+0$ &    $595$ &         $0.1024(20)$ &        $0.06400(76)$ &          $0.560(48)$ & $2.70$ &     $0.0015$ &     $0.0094$ & &     $7802.37086(43)$ &      $4.1877642(29)$ &           $272(99)$ \\
\hline
\end{tabular}
\begin{flushleft}
The fitting uncertainties are given in parenthesis after each estimation, in the units of
the last two figures. Most of the columns have the same meaning as in Table~4 from \citep{Baluev15a}\\
$^1$Defined as $\mathcal Q_{\rm sum}' = \sqrt{\sum_{i=1}^N {Q_i'}^2}$.\\
$^2$The quadratic TTV ephemeris and the value of $\chi^2_{\rm TTV}$ do not include the
\citet{Southworth09a,Southworth09b,Southworth10} DFOSC data, because they may be affected
by clock errors.\\
$^3$Number of lightcurves fitted with individual red noise term + number of lightcurves fitted with shared $\tau$.\\
$^4$Orbital eccentricity of $\sim 0.15$ is not taken into account, see Sect.~\ref{sec_rvttv} for a self-consistent fit.\\
$^5$The realistic uncertainty also depends on the observed TTV scatter $\chi_{\rm TTV}^2$,
likely inspired by the star activity, see text.
\end{flushleft}
\end{table*}

\subsection{WASP-12: a nonlinear TTV trend}
\label{sec_WASP12}
Our analysis yielded $9$-sigma significance of the WASP-12 quadratic TTV term. This appears
convincing, and the trend itself can be easily distinguished in Fig.~\ref{fig_WASP12}
below. We obtained the characteristic orbit decay time $T_{\rm d} = P/|\dot P| = 3.57\pm
0.33$~Myr (or $3.60\pm 0.34$~Myr for the HQ subsample). This is consistent with the recent
estimations by \citet{Patra17} and \citet{Maciejewski18a}. These estimates were based on
the multiplicative noise model \citep{Baluev14a}. The noise scale factor becomes $1.35$ or
$1.33$, the values of $\sqrt{\chi^2_{\rm TTV}}$ from Table~\ref{tab_traw}.

\begin{figure*}
\includegraphics[width=\linewidth]{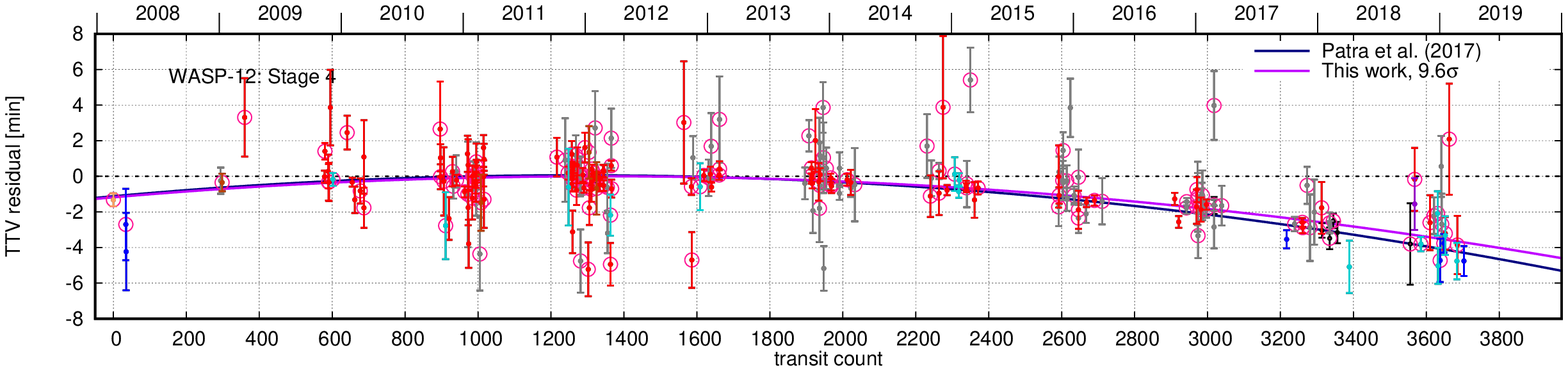}\\
\includegraphics[width=\linewidth]{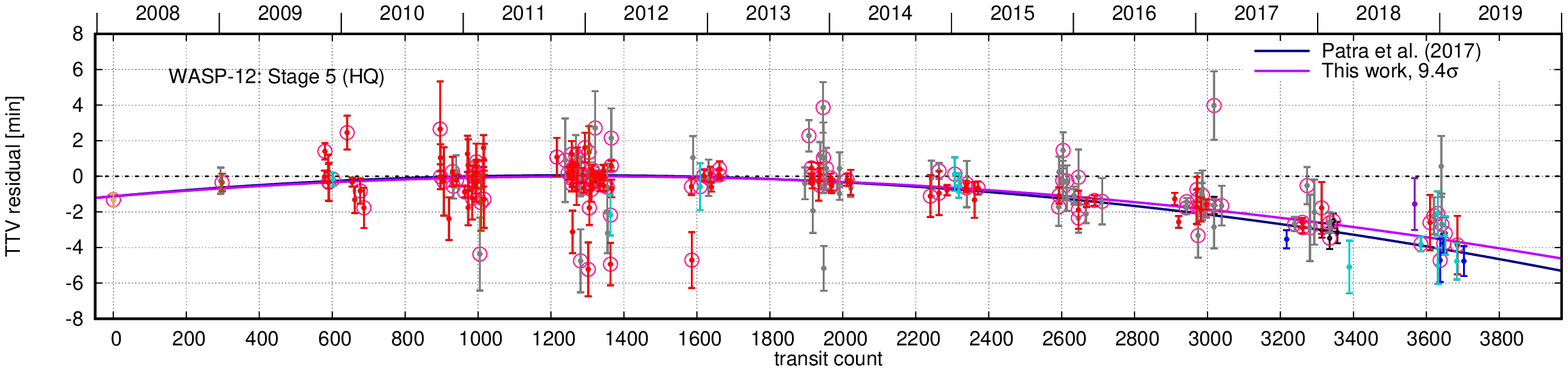}
\caption{Transit times of WASP-12 derived in this work. Top panel is for all the TTV data
(stage~4), bottom panel is for only HQ ones (stage~5). The models of the quadratic TTV
trend are also plotted (for the multiplicative noise model).}
\label{fig_WASP12}
\end{figure*}

We also considered the so-called regularized noise model from \citep{Baluev14a}, which in
our conditions is almost equivalent to the `additive' model. In this model, the noise is
represented as a quadrature sum of the derived TTV uncertainty and of a `jitter'. With this
model we obtain $T_{\rm d} = 3.55\pm 0.31$~Myr ($3.60\pm 0.31$~Myr from only HQ TTVs),
practically the same values. The best fitting TTV jitter for our data is estimated to be
$20.8\pm 2.5$~sec ($18.9\pm 2.5$~sec for the HQ subsample). Therefore, this result is
practically model-invariant and thus very trustable. As such, the tidal quality factor
remains at $Q_* \sim 2 \cdot 10^5$, the value from \citep{Patra17}.

We did not include secondary eclipses in our analysis, and did not use some transit timings
published without lightcurves that could be reprocessed. From only the transit timing data,
we did not obtain any qualitatively new result for WASP-12, but RV data brought a
significant additional information about the nature of this TTV trend (see
Section~\ref{sec_rvttv}).

\subsection{WASP-4: yet another TTV trend?}
\label{sec_WASP4}
We suspected the nonlinear trend in WASP-4, analogous to the WASP-12 one, right after the
new EXPANSION lightcurves from 2017 observing season were processed. The magnitude of the
trend corresponded to $T_{\rm d}\sim 10$~Myr (surprisingly close to what was recently
claimed by \citealt{Bouma19}). However, that time the trend interpretation depended on just
a few data points from 2017. To confirm or retract the trend hypothesis we initiated in
2018 a prioritized observing campaign of WASP-4 within the EXPANSION project. By the end of
2018 we acquired $10$ new transit lightcurves.

Table~\ref{tab_WASP4log} shows the observation log, including the EXPANSION data, as well
as a few older lightcurves found in the ETD and AXA databases, and also $6$ archival
lightcurves from the TRAPPIST-South telescope. This table does not include data taken from
the literature ($29$ lightcurves). The total number of WASP-4 lightcurves reprocessed in
this work was $66$ (plus one outlier not included in the final analysis). The trend
information mainly comes from $14$ observations made in 2017-2018. Among them $10$ were
taken by P.~Evans with a $36$~cm Planewave CDK telescope equipped by a SBIG~STT~1603-3 CCD
and hosted at El Sauce Observatory, Chile. This is a good quality equipment at a good site,
and the corresponding TTV measurements appeared in turn quite competitive with even TESS
ones (which were released later).

\begin{table}
\tiny
\caption{Observation log for $31$ of the WASP-4 transit lightcurves.}
\label{tab_WASP4log}
\begin{tabular}{@{\;}l@{\;}c@{\;}c@{\;}c@{\;}cp{26mm}@{\;}}
\hline
Obs. Date  & Aperture  & Filter    & Cadence & Airmass & Observer \\
           & [m]       &           & [min]   &         &          \\
\hline
2008-09-23 &           &    Rc     & $0.46$   &  $1.26 \longmapsto 1.03$      &   Fernando Tifner (AXA) \\
2009-09-22 &           &    Rc     & $1.18$   &  $1.27 \longmapsto 1.02$      &   Fernando Tifner (AXA) \\
2010-07-08 &           &    Rc     & $1.02$   &  $1.11 \longmapsto 1.19$      &   Thomas Sauer (ETD) \\
2010-08-17 & $2.15$    &    Ic     & $1.13$   &  $1.07 \longmapsto 1.34$      &   Eduardo Fernandez-Lajus, Romina P. Di Sisto \\
2010-10-04 &           &    Clear  & $0.46$   &  $1.02 \longmapsto 1.15$      &   Gavin Milne (ETD) \\
2010-11-01 &           &    Rc     & $0.41$   &  $1.03 \longmapsto 1.38$      &   TG Tan (ETD) \\
2010-11-05 &           &    Clear  & $0.52$   &  $1.13 \longmapsto 1.85$      &   Ivan Curtis (ETD) \\
2010-12-20 & $0.60$    &    `I+z'  & $0.53$   &  $1.14 \longmapsto 2.02$      &   TRAPPIST \\
2011-09-15 & $0.60$    &    `I+z'  & $0.33$   &  $1.62 \longmapsto 1.04$      &   TRAPPIST \\
2011-09-27 & $0.60$    &    Ic     & $0.33$   &  $1.23 \longmapsto 1.05$      &   TRAPPIST \\
2011-10-21 & $0.60$    &    Ic     & $0.33$   &  $1.04 \longmapsto 1.61$      &   TRAPPIST \\
2011-12-19 & $0.60$    &    `I+z`  & $0.33$   &  $1.14 \longmapsto 2.07$      &   TRAPPIST \\
2012-06-07 & $0.60$    &    Ic     & $0.35$   &  $1.41 \longmapsto 1.03$      &   TRAPPIST \\
2012-09-11 & $0.25$    &    Clear  & $0.82$   &  $1.59 \longmapsto 1.14$      &   Phil Evans \\
2013-09-21 &           &    Clear  & $0.79$   &  $1.19 \longmapsto 1.07$      &   Colazo, C. Schneiter, E. M. \\
2013-10-07 &           &    Rc     & $0.30$   &  $1.41 \longmapsto 1.05$      &   Erin Miller (ETD) \\
2013-12-13 &           &    Rc     & $0.31$   &  $1.60 \longmapsto 1.10$      &   Erin Miller (ETD) \\
2014-08-03 & $2.15$    &    Clear  & $3.17$   &  $1.10 \longmapsto 1.04$      &   Eduardo Fernandez-Lajus, Romina P. Di Sisto \\
2014-08-16 &           &    Clear  & $0.64$   &  $1.05 \longmapsto 1.60$      &   Martin Masek (ETD) \\
2014-08-20 & $2.15$    &    Ic     & $6.04$   &  $1.01 \longmapsto 1.27$      &   Eduardo Fernandez-Lajus, Romina P. Di Sisto \\
2014-08-20 & $1.54$    &    Clear  & $3.26$   &  $1.01 \longmapsto 1.42$      &   Carlos Colazo, Carolina Villarreal \\
2014-10-25 & $1.54$    &    Rc     & $0.75$   &  $1.30 \longmapsto 2.91$      &   Cecilia Quinones \\
2015-08-15 & $2.15$    &    Ic     & $3.05$   &  $1.02 \longmapsto 1.22$      &   Eduardo Fernandez-Lajus, Romina P. Di Sisto \\
2017-07-26 & $0.36$    &    Rc     & $1.51$   &  $1.42 \longmapsto 1.04$      &   Phil Evans \\
2017-09-07 & $0.36$    &    Rc     & $1.19$   &  $2.44 \longmapsto 1.02$      &   Phil Evans \\
2017-09-23 & $0.36$    &    Rc     & $2.16$   &  $1.38 \longmapsto 1.02$      &   Phil Evans \\
2017-09-24 & $0.46$    &    Clear  & $1.02$   &  $1.19 \longmapsto 1.15$      &   H. Durantini Luca, P. Baez, C. Colazo \\
2018-05-23 & $0.36$    &    Rc     & $2.07$   &  $2.74 \longmapsto 1.09$      &   Phil Evans \\
2018-06-20 & $0.36$    &    Rc     & $2.03$   &  $1.05 \longmapsto 1.02$      &   Phil Evans \\
2018-07-25 & $0.36$    &    Rc     & $2.26$   &  $1.75 \longmapsto 1.02$      &   Phil Evans \\
2018-08-10 & $0.36$    &    Rc     & $2.06$   &  $1.08 \longmapsto 1.31$      &   Phil Evans \\
2018-08-12 & $0.30$    &    Rc     & $1.55$   &  $1.10 \longmapsto 1.30$      &   Carl R. Knight \\
2018-08-14 & $0.36$    &    Rc     & $2.22$   &  $1.03 \longmapsto 1.32$      &   Phil Evans \\
2018-08-15 & $2.15$    &    Ic     & $3.05$   &  $1.05 \longmapsto 1.28$      &   Eduardo Fernandez-Lajus, Romina P. Di Sisto \\
2018-08-22 & $0.36$    &    Rc     & $2.12$   &  $1.02 \longmapsto 1.51$      &   Phil Evans \\
2018-08-26 & $0.36$    &    Rc     & $2.06$   &  $1.02 \longmapsto 1.58$      &   Phil Evans \\
2018-10-14 & $0.30$    &    Rc     & $1.61$   &  $1.10 \longmapsto 1.29$      &   Carl R. Knight \\
\hline
\end{tabular}
\end{table}

Our new data did not confirm the trend: the updated TTV time series became consistent with
strictly linear ephemeris, so we decided that our trend hypothesis was wrong. But
\citet{Bouma19} reported a detection of this trend based on the new TESS transit data,
obtained practically simultaneously with our observations in the EXPANSION network. To shed
more light on this apparent controversy, we then performed additional analysis, including
the TTV data published in the literature without lightcurves and the new TESS timings. This
includes very accurate transit times derived from the transmission spectroscopy by
\citet{Huitson17}, transit times by \citet{Hoyer13}, by \citet{Wilson08} and two early WASP
timings given in \citep{Gillon09a}. We did not use the HST spectral observations from
\citet{Ranjan14}: these data might be inaccurate because the spectra were partly
overexposed and hence the flux measurements are likely not very reliable.

\begin{figure*}
\includegraphics[width=\linewidth]{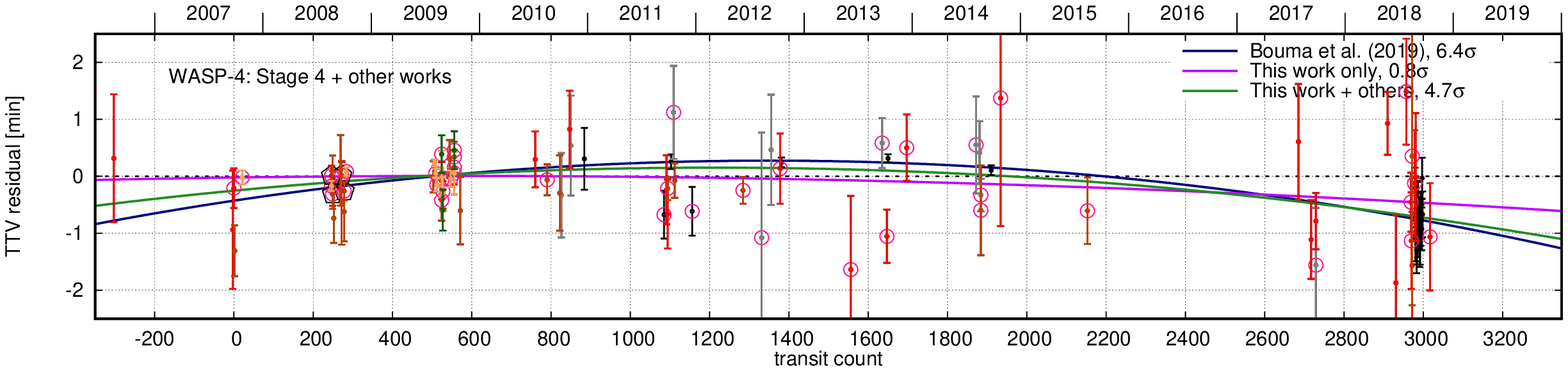}\\
\includegraphics[width=\linewidth]{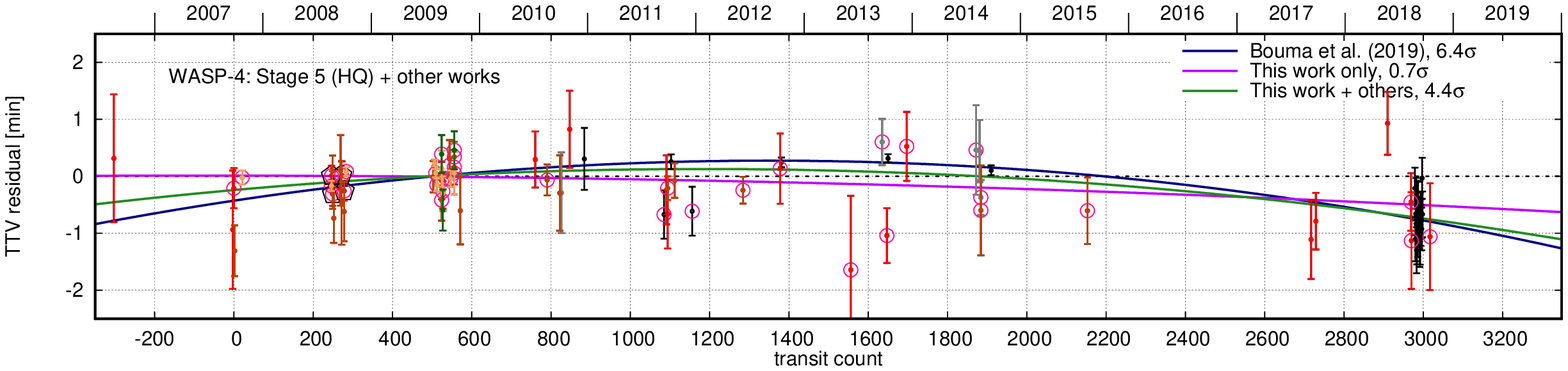}
\caption{Transit times of WASP-4, including the homogeneous sample from this work (without
DFOSC data possibly affected by clock errors), and the timing data published in literature
without lightcurves. Top panel is for all the TTV data (Stage~4), bottom panel is for only
HQ ones (Stage~5). Several models of the quadratic TTV trend are also plotted with the
trend significance labelled in the legend (for the regularized noise model and simply
merging heterogeneous TTV data).}
\label{fig_WASP4}
\end{figure*}

The full TTV time series is shown in Fig.~\ref{fig_WASP4}. Now, with the new TESS transit
times added, the quadratic term of the trend indeed appears significant, according to our
analysis. However, we obtain a smaller magnitude and significance than \citet{Bouma19}
reported. The trend is still not detectable with the use of only the homogeneously derived
portion of TTV data from this work. That is, the information about the trend comes mainly
from the third-party observations rather than from our data release. By inspecting
Fig.~\ref{fig_WASP4} we may suspect that the trend depends primarily on just the four
high-accuracy timings provided by \citet{Huitson17}. The TESS timing does not in fact
contradict anything and visually they are in a satisfactory agreement with what was
obtained in the EXPANSION project in 2018.

However, justifying the trend detection based on just four data points, even apparently
accurate ones, might be quite dangerous. Looking into the details of the \citep{Huitson17}
TTV data, they were based on just the \emph{linear} limb-darkening model. Although the
authors ensured that based on some preliminary analysis their results (including fit
uncertainties) did not change significantly for linear and for more complicated
limb-darkening models, we remain concerned about this. Also, we could not find a clear
confirmation in the text that the red noise was taken into account when fitting the
lightcurves. Although it is mentioned that some `systematics' are fitted, from the
description given in the text the `systematics' appear to be a deterministic parametric
function rather than an autocorrelated random process.

In view of this we notice that in the similar transmission spectroscopy lightcurves for
WASP-12 \citep{Stevenson14} we robustly detect significant red noise. Inclusion of this red
noise in the lightcurve model roughly doubled the derived transit timing uncertainties from
$\sim 3$~sec to $\sim 5$~sec. Significant red noise was also detected in the WASP-52
transmission spectroscopy lightcurve from \citep{Chen17}, though not detected in the
HD~189733 data by \citet{Kasper19}. The latter, however, revealed the anomalous transit
time discussed above. A public release of the \citet{Huitson17} lightcurves is not
available, so we did not reanalyse them in our pipeline. We therefore decided to
investigate this issue using a different approach.

As it was explained above, formally declared TTV uncertainties never appear entirely
accurate: the actual scatter of TTV residuals may be systematically different (usually
larger). However, different teams may process data quite differently, and hence each team
might have its own bias in the reported TTV uncertainties. Therefore, different portions of
such a heterogeneous TTV compilation may need to be weighted differently to balance this
effect. However, those weights are not known to us a priori, so they need to be estimated
from the TTV data `on-the-fly', e.g. based on the actually observed scatter of the TTV
residuals in each homogeneous portion.

We therefore separated all our TTV data into the following four more or less homogeneous
classes: (i) the `main' subset including transit timings derived in this work and three old
timings given in \citep{Gillon09a} without public lightcurve data; (ii) the rich TTV subset
by \citet{Hoyer13}; (iii) the four high-accuracy timings by \citet{Huitson17}; and (iv) the
TESS timings from \citep{Bouma19}. All these data sets should have an independently
fittable noise parameter.

This noise was modelled by one of two models discussed in \citep{Baluev14a}, namely by (i)
the multiplicative model, or (ii) the so-called regularized model. These `noise models'
represent a parametrized model for the variance of each TTV measurement, in which a single
free parameter regulates the weight of the corresponding TTV data set as a whole. Since
this approach involves a separate and largely independent treatment of each TTV data set,
we call this as `separated' model of the TTV data. It can be fitted by using the
maximum-likelihood method, as discussed in \citep{Baluev08b}. In such a way the relative
weighting of different TTV subsets is determined adaptively and basically tied to the
corresponding TTV residuals RMS.

For a comparison, we also analysed the TTV data plainly merged into a single time series
without any relative weighting. This analysis was also performed for the same two noise
models, multiplicative and regularized ones. The TTV trend itself was always modelled by
the quadratic function~(\ref{trend}) with three free coefficients.

As we expected, it appeared that the magnitude of the quadratic term and especially its
derived uncertainty is sensitive to the choice of the noise model. In the case of a
`separated' model the trend uncertainty gets increased. Therefore, by allowing some TTV
data to be actually less accurate than stated, the significance of the trend may reduce.
For example, it may reduce if the four \citet{Huitson17} transit times are less accurate
than formally stated. And because of the small number of these data (just the four), their
RMS does not constrain the noise level well, so this level can be varied relatively freely.

\begin{figure*}
\includegraphics[width=0.33\linewidth]{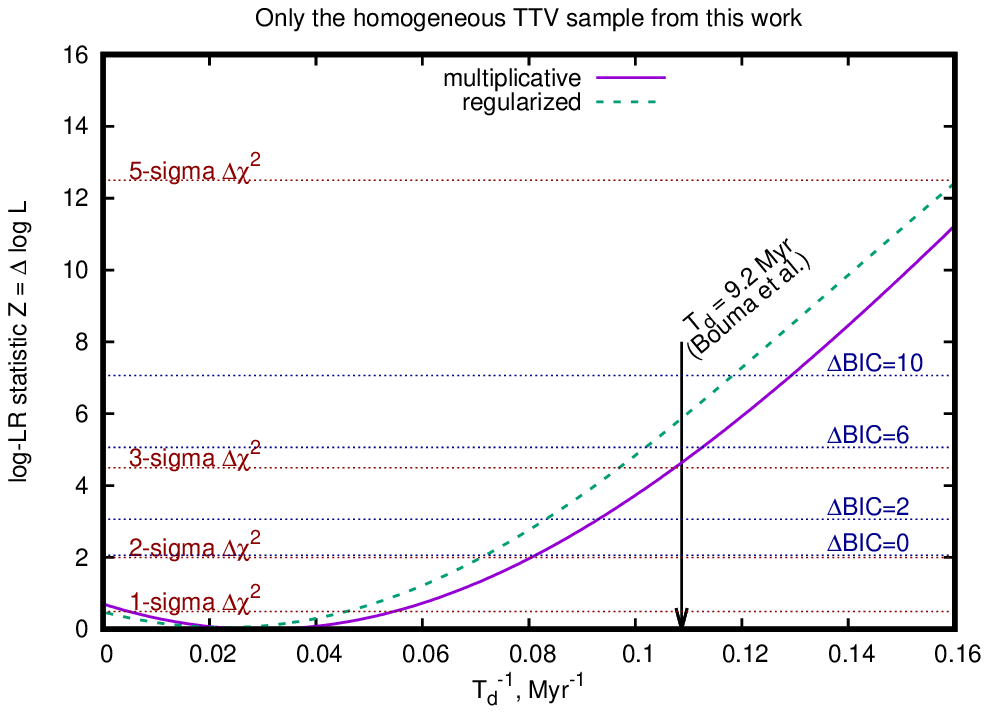}
\includegraphics[width=0.33\linewidth]{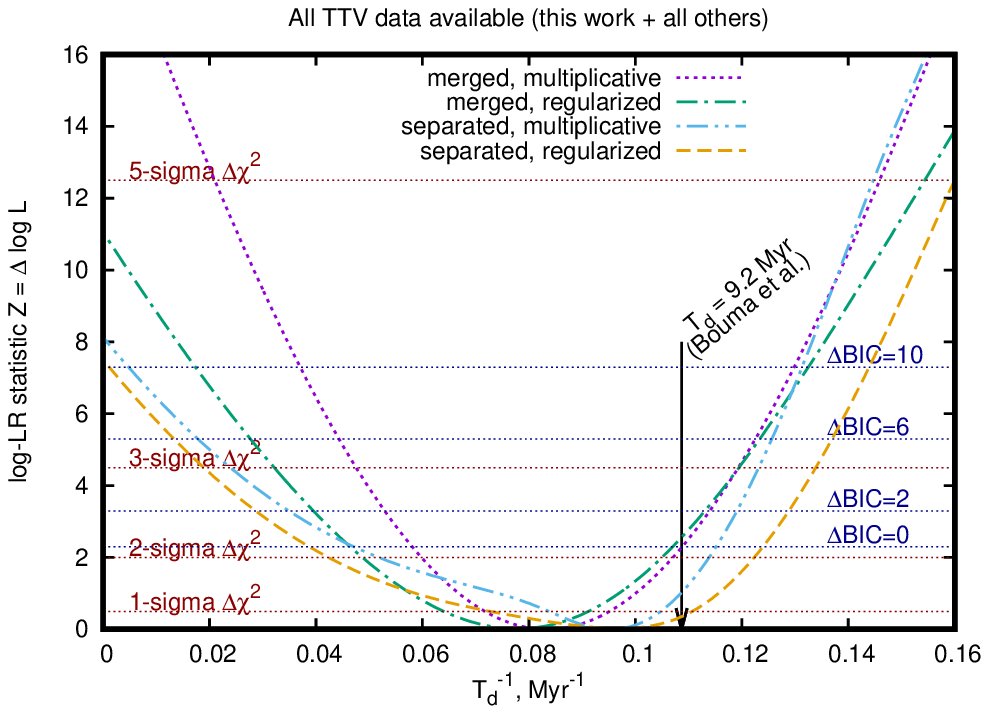}
\includegraphics[width=0.33\linewidth]{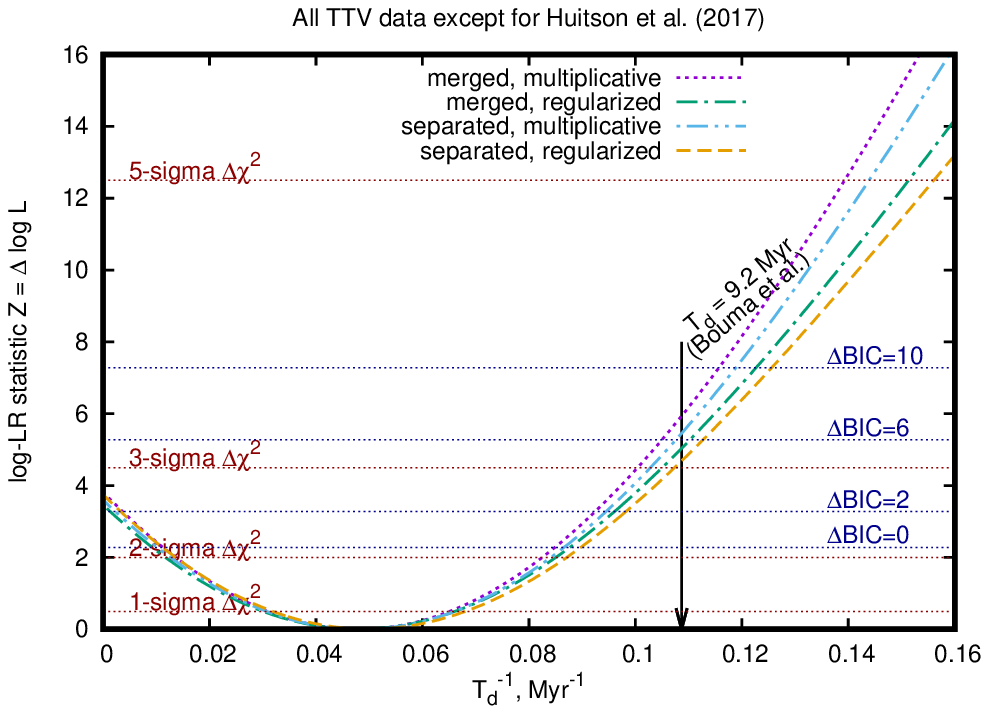}
\caption{Logarithm of the likelihood ratio statistic for WASP-4, $Z(q)$, as a function of
$q = T_{\rm d}^{-1}$. Three graphs correspond to different compilations of TTV data
(homogeneous from this work / all / all but \citet{Huitson17}). The curves within each
graph correspond to different models of the TTV noise (merged/separated and
multiplicative/regularized). In each graph a set of the significance threshold levels is
also shown, corresponding to the frequentist $\chi^2$ test or to the Bayesian information
criterion (BIC). See text for more details.}
\label{fig_WASP4-lik}
\end{figure*}

In Fig.~\ref{fig_WASP4-lik}, we demonstrate this effect in the shape of the likelihood
function $\mathcal L$. For this goal we consider the log-likelihood-ratio statistic $Z$
determined in accordance with \citep{Baluev08b}. We compute (i) the global maximum of the
likelihood function $\mathcal L_{\max}$ with respect to all the noise parameters and all
three TTV trend coefficients, and (ii) the value $\mathcal L_{\max}'(q)$ maximized with
respect to all parameters except for the quadratic coefficient $dP/dn=P^2 T_{\rm d}^{-1} =
q P^2$, where $q=T_{\rm d}^{-1}$ was fixed prior to the fit. The quantity $Z(q) =
\log[\mathcal L_{\max}/\mathcal L_{\max}'(q)]$ therefore indicates whether the given $q$ is
statistically consistent with the best-fitting value $\hat q$ which corresponds to the
global maximum $\mathcal L_{\max}$. We always have $Z(q)\geq 0$, and the larger is $Z$, the
more statistically significant is the deviation of $q$ from $\hat q$ and the less
consistent with the data this $q$ is. If our models are linearisable than $Z(q)$ should
have an almost parabolic shape with a single minimum at $\hat q$.

We use two approaches to calibrate the levels of $Z(q)$, both rely on the assumption that
the model is linearisable and hence $Z(q)$ is quadratic (while the likelihood ratio
$\exp(-Z(q))$ is Gaussian). The first approach is the frequentist $\chi^2$ test, and the
second one is the Bayesian Information Criterion (BIC). In the frequentist treatment, the
significance level of a given $Z$ is approximately the $\chi^2_1(2Z)$, or the
$\chi^2$-distribution with $d=1$ degree of reedom (one degree because we have just one free
parameter $q$ left in $Z(q)$). This would mean that the significance level for a given $q$
would correspond to $\sqrt{2Z}$ in the $n$-sigma notation, or vice versa, any $n$-sigma
significance level would correspond to the threshold level $Z(q)=n^2/2$.

The BIC is defined as ${\rm BIC} = 2\log\mathcal L - k \log N$, where $k$ is the total
number of free parameters in the model, and $N$ is the number of observations (number of
transit times). To compare different models with $k_1$ and $k_2$ parameters we use the
difference $\Delta{\rm BIC} = 2 Z - d \log N$ with $d=k_1-k_2=1$ in our case. Hence, the
significance threshold for $Z(q)$ becomes $Z = (\Delta {\rm BIC} + \log N)/2$. Here $\Delta
{\rm BIC}$ is deemed to be an input parameter determining the requested significance level
(typical practical values are $2$, $4$, $6$, $10$).

The special value $Z(0)$ indicates the significance of the nonlinear trend itself (i.e. how
much $q=0$ is consistent with the data, with the adopted TTV noise model).

In Fig.~\ref{fig_WASP4-lik}, we plot this statistic $Z(q)$ for three TTV data compilations,
including (i) only the homogeneous data from this work, (ii) all TTV data, (iii) all TTV
data excluding \citep{Huitson17}, and for all our noise models, including (i) the plain
merging of heterogeneous datasets and (ii) adaptive merging of heterogeneous datasets with
individually fittable noise parameters. For each of these model layouts we adopt either a
multiplicative or regularized noise model, defined in \citep{Baluev14a}.

As we can see, the shape of the likelihood function may change a lot depending on the model
and TTV data involved. We can draw the following conclusions:
\begin{enumerate}
\item Our homogeneously derived TTV data do not support the existence of any quadratic
trend. These data are consistent with a linear ephemeris below $1$-sigma level.

\item Simultaneously, the value of $T_{\rm d}=9.2$~Myr from \citep{Bouma19} seems too
poorly consistent with our homogeneous TTV subsample, at the level above $3$-sigma in terms
of the $\chi^2$ test or with $\Delta {\rm BIC}\sim 4-6$ (depending on the model). We
believe this may appear, at least in some part, because \citet{Bouma19} did not take into
account the heterogeneous nature of the TTV data, merging them into a single time series.

\item Joining our data with the remaining third-party TTV measurements allows us to refine
the localization of the parameter $q$ greatly and even suggests that this $q$ can be
significantly non-zero. However, the significance of this conclusion, as well as possible
confidence ranges for $q$ appear very model dependent. If we plainly merge all the TTV
data, we obtain that $q$ is inconsistent with zero at the high $\sim 5-7$-sigma level. But
using our adaptive separated noise model, this significance drops to merely $3-4$-sigma.

\item The shape of the likelihood function becomes significantly non-parabolic in the case
of our adaptive separated noise model. This indicates that this model may be too non-linear
and therefore our significance estimates may appear inaccurate. It may even appear that the
significance of the trend is reduced even further below the $3-4$-sigma level mentioned
above.

\item The most trustable and model-stable behaviour appears when we just remove the TTV
data by \citet{Huitson17}. Then $Z(q)$ behaves as a nice parabolic function, indicating an
almost-linear model and nearly Gaussian likelihood. In this case, the quadratic trend has
the significance $2.8-3$-sigma or $\Delta {\rm BIC} \sim 4$, which is very remarkable but
still needs further confirmation by more observations. The magnitude of the best fitting
trend then becomes $T_{\rm d} \sim 20$~Myr with large uncertainty. Curiously, the value of
$T_{\rm d}=9.2$~Myr given by \citet{Bouma19} appears in this case even less likely than the
no-trend model ($T_{\rm d}=\infty$).

\item In any case, the trend magnitude is very uncertain, while its confidence ranges
appear very asymmetric and non-Gaussian in the separated noise model. The value $T_{\rm
d}=9.2$~Myr given in \citep{Bouma19} looks more like a lower limit on $T_{\rm d}$, while
the actual value may reach even $\sim 100$~Myr, given the large uncertainty of this
parameter.
\end{enumerate}

Therefore the putative TTV trend magnitude and the detection significance for WASP-4 are
severely model-dependent. They solely depend on how we treat the heterogeneous nature of
the TTV data. Moreover, as recognized by \citet{Bouma19}, $T_{\rm d}$ as small as $9.2$~Myr
is inconsistent with theoretical predictions of the tidal quality parameter. Given our
discussion, we believe that it is too early to definitely claim the detection of this trend
until more homogeneous TTV data are collected. At least, it is too early to claim that this
object breaks any theoretical predictions. However, WASP-4 remains a very interesting
target that may indeed hide serendipitous discoveries.

\section{Self-consistent analysis of radial velocity and transit data}
\label{sec_rvttv}
Below we layout our goals related for the self-consistent analysis of radial velocity and
transit data.
\begin{enumerate}
\item Derive a more complete set of parameters in a self-consistent model, in particular
planetary masses and physical radii (rather than merely the planet/star radii ratio).
\item Derive a more realistic fit of GJ~436~b, taking into account its significant orbital
eccentricity.
\item For WASP-12 and WASP-4, test whether their (possible) TTV trends could appear through
the light-travel effect, caused by the gravity of a distant unseen companion.
\item Derive the rotation parameters of the stars via the Rossiter-McLaughlin (hereafter
RM) effect, and test how much it is sensitive to the correction coeffiecients suggested in
\citep{BaluevShaidulin15}.
\end{enumerate}

Notice that even the combination of transit and radial velocity data does not allow to
determine the star mass from a self-consistent fit. The information about the star mass
usually comes from astrophysical models of stellar spectra, e.g. based on stellar
evolutionary tracks. Such models in fact provide certain constraints on the stellar mass
$M_\star$ and radius $R_\star$ that can be used to provide an entirely self-consistent
global fit. However, in this work we were more interested to estimate the uncertainties
inferred by the transit and radial velocity data, so we still prefer not to mix them with
the uncertainties of astrophysical models that may also contain an additional systematic
error.

\begin{table}
\caption{Star masses adopted in the joint transit+RV fits.}
\label{tab_masses}
\begin{tabular}{lll}
\hline
host star    & $M_\star [M_\odot]$ & reference \\
\hline
Corot-2      & $0.97(6)$           & \citet{Alonso08}    \\
GJ436        & $0.452\left(+14\atop-12\right)$& \citet{Torres08}\\
HAT-P-13     & $1.22\left(+5\atop-10\right)$  & \citet{Bakos09}\\
HD189733     & $0.806(48)$         & \citet{Torres08}\\
TrES-1       & $0.878\left(+38\atop-40\right)$& \citet{Torres08}\\
WASP-2       & $0.84\left(+11\atop-12\right)$ & \citet{Triaud10}\\
WASP-3       & $1.24\left(+6\atop-11\right)$  & \citet{Pollaco08}\\
WASP-4       & $0.930\left(+54\atop-53\right)$& \citet{Triaud10}\\
WASP-5       & $1.000\left(+63\atop-64\right)$& \citet{Triaud10}\\
WASP-6       & $0.88\left(+5\atop-8\right)$   & \citet{Gillon09b}\\
WASP-12      & $1.434\left(+110\atop-90\right)$&\citet{Collins17}\\
XO-2N        & $0.96(5)$           & \citet{Damasso15}\\
XO-5         & $0.88(3)$           & \citet{Pal08}\\
\hline
\end{tabular}
\end{table}

Therefore, we fixed certain `reference' values of $M_\star$ for our ten targets, as given
in Table~\ref{tab_masses}. We did not take into account the stated uncertainties of
$M_\star$ when computing our fits. In case if the adopted $M_\star$ is different from the
reference value, the fit can be easily rebased to another $M_\star'$ based on the following
simple laws:
\begin{eqnarray}
R_\star' &=& R_\star \left(M_\star'/M_\star\right)^{\frac{1}{3}}, \nonumber\\
r_{\rm pl}' &=& r_{\rm pl} \left(M_\star'/M_\star\right)^{\frac{1}{3}}, \nonumber\\
a' &=& a \left(M_\star'/M_\star\right)^{\frac{2}{3}}, \nonumber\\
m_{\rm pl}' \sin i' &=& m_{\rm pl} \sin i \left(M_\star'/M_\star\right)^{\frac{1}{3}}, \nonumber\\
\cos i'  &=& \cos i \left(M_\star'/M_\star\right)^{-\frac{1}{3}}.
\label{scl}
\end{eqnarray}

The first formula of this list comes from the known property that a transit fit actually
constrains the star density $\rho_\star \propto M_\star R_\star^{-3}$, rather than
$R_\star$ or $M_\star$ separately \citep{MandelAgol02}. The second one appears because the
transit data constrain the ratio $r_{\rm pl}/R_\star$, so the scaling law of $r_{\rm pl}$
is the same as for $R_\star$. The third and the fourth formulae for the orbital semimajor
axis and planetary mass, respectively, follow from the basic properties of the Doppler
method and can be found e.g. in \citep{Baluev13c}. The last relationship for cosine of
orbital inclination $i$ follows because $i$ is constrained by only the transit data, via
the measured impact parameter $b = a \cos i/R_\star$, so the scale law for $\cos i$
corresponds to $R_\star/a$. The last two formulae can be combined together to obtain
\begin{equation}
m_{\rm pl}'  = m_{\rm pl} \frac{\left(M_\star'/M_\star\right)^{\frac{1}{3}}}{\sqrt{1 + \left(1-\left(M_\star'/M_\star\right)^{-\frac{2}{3}}\right) \cotan^2 i}}.
\label{mscl}
\end{equation}
Since $\cotan i$ is below $0.15$ for all our targets, the term beneath the square root
represents only a negligible correction. Some scale formulae above are not entirely
accurate, neglecting certain second-order corrections, but since the values of $M_\star$
for all our targets are now restricted to quite narrow ranges, the practical accuracy
of~(\ref{scl},\ref{mscl}) should be satisfactory.

To perform the self-consistent analysis, for the transit data we used basically the same
model as above, except that the planet motion was assumed Keplerian rather than circular.
The radial velocity for each target was modelled by the Keplerian curve plus a linear trend
(to account e.g. for possible long-period companions in the system). We also include a
quadratic term in the planetary longitude, to take into account possible TTV trends, see
\citep{Baluev18c}.

For many of our targets, the RV data contained substantial in-transit runs, obviously aimed
to detect the RM effect. For these targets we therefore included in our compound RV model
the RM effect based on the approach by \citet{BaluevShaidulin15}. To accurately approximate
this effect, we must know some effective values of the limb-darkening coefficients $A_{\rm
RV}$ and $B_{\rm RV}$, corresponding to the Doppler spectral range. Also, we need to
specify two correction coefficients $\nu$ and $\mu$ that depend on the average
characteristics of spectral lines and on the method used to derive the radial velocity from
the spectrum. But unfortunately, these four quantities are too difficult to derive reliably
from the spectra themselves. Instead, it is reasonable to treat them as fittable parameters
of the RV model. However, in such a case the model becomes nearly degenerate, because as
discussed in \citep{BaluevShaidulin15}, the parameters $\nu$ and $\mu$ are strongly
correlated with $A$ and $B$. We therefore adopted the following hybrid approach. First, we
assumed that $A_{\rm RV}$ and $B_{\rm RV}$ are equal to the corresponding values of the
photometric data obtained with a clear aperture. Concerning $\nu$ and $\mu$, we considered
them separately for different instruments. This should take into account possible
corrections of the RM effect, jointly with possible inaccuracies of the limb-darkening
coefficients, on a per-instrument basis.

When computing the fits, we treated separately the RV data obtained at different
instruments. Moreover, if the RV data belonging to the same instrument contained in-transit
pieces, we separated these portions of the data from each other and treated them as
individual RV data sets. This might make the model more adequate, because e.g. the scatter
of the RV residuals within each such short run covering just a few hours is significantly
smaller than for the entire dataset covering years. This is essentially the impact of red
noise in the RV data. Also, the red noise may result in a small individual offset of each
in-transit run.

For some targets we also found several compact series of out-of-transit runs covering just
a single night. Such portions of the RV data were treated separately too, with an
individual offset and individual noise parameters. After performing a preliminary fit of
the joint transit+RV model described above, we run the red-noise detection algorithm
described above for transit data, but now we extended it to all the RV data sets as well.

For Corot-2, WASP-4, and WASP-12 we computed an additional alternative fit without
splitting the RV data belonging to the same instrument. In this case, possible RV offsets
between different compact runs were taken into account implicitly, via a single red noise
model of the merged data set. All the analysis was performed with the PlanetPack software
of version~3 \citep{Baluev18c}.

We separate our results in two parts: Table~\ref{tab_scs} gives some most important
non-planetary parameters of our fits, and Table~\ref{tab_scp} contains only planetary
parameters. The tables are presented here in a reduced form; their expanded versions that
include e.g. RM correction coefficients can be found in the online-only supplement.

First of all, we notice that the RM correction coefficients $\nu$ and $\mu$ are usually
consistent with zero, given their uncertainties. We found only the following targets
convincingly demonstrating significant nonzero values: Corot-2 ($\mu$ for the HARPS RV
data), HD~189733 ($\nu$ for Keck/HIRES and SOPHIE, and $\mu$ for HARPS, HARPSN, and
Keck/HIRES), WASP-5 ($\mu$ for HARPS), and possibly GJ~436 ($\nu$ for GJ~436 HARPS, HARPSN,
Keck/HIRES). In theory, $\nu$ should usually be zero, since most our RV data were derived
with TERRA, which is a kind of a spectrum modelling method \citep{AngladaEscudeButler12}. A
nonzero $\nu$ may appear only if the RV data were derived by the cross-correlation
technique, and simultaneously the spectral lines have some asymmetry on average
\citep{BaluevShaidulin15}. Since this is not the case for the most targets, a few
significantly nonzero $\nu$ estimates may indicate that the adopted limb-darkening
coefficients $A_{\rm RV}$ and $B_{\rm RV}$ are inaccurate for the relevant RV dataset. In
such a case, the value of $v\sin i$ and $\lambda$ may involve an additional bias which is
difficult to estimate without a better guess for $A_{\rm RV}$ and $B_{\rm RV}$.

Even though we did not detect here very many occurrences of significantly nonzero RM
correction coefficients, it is still important to preserve them in the RV model as free
parameters, in order to have a more realistic (increased) uncertainty in $v\sin i$ and
$\lambda$, as well as in other RV-derived parameters, like e.g. the RV trends.

\begin{table}
\tiny
\caption{Self-consistent fits of transit and radial velocity data: stellar parameters and RV trends.}
\label{tab_scs}
\begin{tabular}{lllllll}
\hline
host star   & rad. accel.       & star radius         & rotation vel.   & spin-orbit ang.       \\
            & $c_1$~[m/s/yr]    & $R_\star\ [R_\odot]$& $v\sin i$~[m/s] & $\lambda\ [^\circ]$   \\
\hline
Corot-2     & $-74(11)$         & $0.925(19)$         & $8930(400)$     & $3.1(6.6)$            \\
Corot-2'    & $-0.5(3.1)$       & $0.885(29)$         & $8570(430)$     & $3.0(8.2)$            \\
GJ436       & $0.02(11)$        & $0.4266(89)$        & $1800(1000)$    & $354.5(4.9)$          \\
HAT-P-13    & $17.2(1.0)$       & $1.691(25)$         & $2300(1100)$    & $359.5(5.5)$          \\
HAT-P-13'   & $17.56(85)$       & $1.692(25)$         & $2600(1300)$    & $0.0(5.4)$            \\
HD189733    & $-1.07(75)$       & $0.7583(31)$        & $2280(110)$     & $0.08(36)$            \\
TrES-1      & $-1.9(1.4)$       & $0.819(13)$         & ---             & ---                   \\
WASP-2      & $-3.9(1.8)$       & $0.8200(82)$        & $960(850)$      & $345(18)$             \\
WASP-3      & $-7.9(2.7)$       & $1.337(28)$         & ---             & ---                   \\
WASP-4      & $-0.7(1.1)$       & $0.9029(43)$        & $1890(220)$     & $344(16)$             \\
WASP-4'     & $-0.8(1.2)$       & $0.9045(46)$        & $1890(220)$     & $343(16)$             \\
WASP-5      & $-0.03(78)$       & $1.127(16)$         & $2470(400)$     & $0.2(6.4)$            \\
WASP-6      & $-420(260)$       & $0.829(23)$         & $1620(130)$     & $352(11)$             \\
WASP-12     & $-5.4(2.0)$       & $1.657(12)$         & $600(1200)$     & $9(48)$               \\
WASP-12'    & $-7.5(2.2)$       & $1.660(16)$         & $1300(1500)$    & $75(30)$              \\
XO-2N       & $-1.2(1.2)$       & $0.986(14)$         & ---             & ---                   \\
XO-5        & $1.3(1.2)$        & $1.068(51)$         & ---             & ---                   \\
\hline
\end{tabular}\\
The fitting uncertainties are given in parenthesis after each estimation, in the units of
the last few figures. The star mass from Table~\ref{tab_masses} was assumed constant here
and its uncertainties were not included in the fit. Using the full set of transits for
GJ436, while only the HQ (Stage 5) transits for other targets. A stroke stands for an
alternative fit computed without splitting the RV data belonging to the same instrument
(see text).
\end{table}

\begin{table*}
\tiny
\caption{Self-consistent fits of transit and radial velocity data: planetary parameters.}
\label{tab_scp}
\begin{tabular}{llllllllll}
\hline
host star    & planet mass                & planet radius              & orbital period$^1$   & TTV trend$^{1,2,3}$&mean longitude$^1$& inclination  & eccentricity   & pericenter arg.     \\
             & $m_{\rm pl}\ [M_{\rm Jup}]$& $r_{\rm pl}\ [R_{\rm Jup}]$& $P$~[d]              & $T_{\rm d}$~[Myr]  & $l\ [^\circ]$  & $i\ [^\circ]$  & $e$            & $\omega\ [^\circ]$  \\
\hline
Corot-2      & $2.744(51)$                & $1.523(36)$                &     $1.74299686(44)$ & $-11.2(9.4)$       & $359.4(1.6)$   & $88.15(56)$    & $0.042(18)$    & $60(18)$            \\
Corot-2'     & $2.85(10)$                 & $1.455(51)$                &     $1.74299682(45)$ & $-10.4(8.2)$       & $0.7(2.4)$     & $88.32(52)$    & $0.013(26)$    & $320(130)$          \\
GJ436        & $0.06896(56)$              & $0.3581(97)$               &     $2.64389856(36)$ & $7.1(3.2)$         & $335.82(44)$   & $86.83(10)$    & $0.1666(57)$   & $324.9(2.4)$        \\
HAT-P-13 b   & $0.8528(59)$               & $1.465(30)$                &     $2.9162384(17)$  & $-2.42(86)$        & $158.64(41)$   & $82.12(22)$    & $0.0126(46)$   & $219(29)$           \\
\phantom{HAT-P-13}
c            & $14.17(28)$                & ---                        &     $446.32(24)$     & ---                & $63.99(18)$    & ---            & $0.6621(58)$   & $175.28(37)$        \\
HAT-P-13' b  & $0.8532(53)$               & $1.466(30)$                &     $2.9162383(17)$  & $-2.41(85)$        & $158.58(35)$   & $82.10(22)$    & $0.0117(41)$   & $218(28)$           \\
\phantom{HAT-P-13'}
c            & $14.16(25)$                & ---                        &     $446.29(21)$     & ---                & $63.99(16)$    & ---            & $0.6614(50)$   & $175.28(32)$        \\
HD189733     & $1.1542(74)$               & $1.1840(52)$               &    $2.218575123(57)$ & $28(12)$           & $20.45(38)$    & $85.712(36)$   & $0.0028(38)$   & $62(64)$            \\
TrES-1       & $0.6967(82)$               & $1.122(22)$                &     $3.03006960(18)$ & $-5.7(9.5)$        & $298.67(49)$   & $88.69(28)$    & $0.003(12)$    & $263(86)$           \\
WASP-2       & $0.8711(73)$               & $1.087(14)$                &     $2.15222163(67)$ & $-12(11)$          & $214.37(25)$   & $84.82(10)$    & $0.0134(56)$   & $253(11)$           \\
WASP-3       & $1.982(49)$                & $1.419(32)$                &     $1.84683480(30)$ & $-11.1(5.5)$       & $274.88(83)$   & $84.24(32)$    & $0.010(15)$    & $41(74)$            \\
WASP-4       & $1.1949(65)$               & $1.3915(82)$               &    $1.338231501(75)$ & $47(45)$           & $235.80(26)$   & $88.63(30)$    & $0.0068(35)$   & $258(20)$           \\
WASP-4'      & $1.1976(68)$               & $1.3940(86)$               &    $1.338231501(75)$ & $47(45)$           & $235.87(26)$   & $88.63(30)$    & $0.0053(38)$   & $247(28)$           \\
WASP-5       & $1.5351(80)$               & $1.294(25)$                &     $1.6284311(14)$  & $5.2(6.3)$         & $343.01(32)$   & $84.57(33)$    & $0.0086(46)$   & $66(22)$            \\
WASP-6       & $0.458(20)$                & $1.175(35)$                &     $3.36100264(65)$ & $14(14)$           & $32.6(2.4)$    & $89.00(36)$    & $0.036(24)$    & $116(36)$           \\
WASP-12      & $1.422(14)$                & $1.953(15)$                &    $1.091421080(96)$ & $3.46(24)$         & $37.00(51)$    & $81.86(16)$    & $0.0259(74)$   & $250(11)$           \\
WASP-12'     & $1.413(15)$                & $1.956(20)$                &    $1.091421078(96)$ & $3.47(24)$         & $36.82(49)$    & $81.96(18)$    & $0.024(11)$    & $252(11)$           \\
XO-2N        & $0.5924(68)$               & $1.017(16)$                &     $2.61585963(16)$ & $80(140)$          & $198.72(48)$   & $88.33(25)$    & $0.008(13)$    & $91(29)$            \\
XO-5         & $1.050(15)$                & $1.061(67)$                &     $4.1877477(38)$  & $-1.82(68)$        & $83.5(1.2)$    & $86.82(51)$    & $0.009(12)$    & $200(94)$           \\
\hline
\end{tabular}
\begin{flushleft}
Same comments as in Table~\ref{tab_scs} also apply here.\\
$^1$These parameters refer to $T_0=2455197.5$ (1~Jan, 2015) in the BJD TDB system.\\
$^2$Except for WASP-12 case, the uncertainties for $T_{\rm d}$ are rather formal here,
because this parameter becomes very nonlinear and non-Gaussian whenever it is comparable to
the uncertainty. The linear parameter is $q=1/T_{\rm d}=-\dot P/P$ with the uncertainty
$\sigma_q=\sigma_{T_{\rm d}}/T_{\rm d}^2$.\\
$^3$The realistic uncertainty in $T_{\rm d}$ also depends on the observed TTV scatter
$\chi_{\rm TTV}^2$ from Table~\ref{tab_traw}, see Section~\ref{sec_traw}.
\end{flushleft}
\end{table*}

Now let us consider the putative RV trends (radial accelerations). They are modelled by a
linear function
\begin{equation}
{\rm RV} = c_0' + c_1 (t-t_0),
\label{rvtrend}
\end{equation}
where $c_0'$ is a dataset-specific RV offset, and $c_1$ is the trend coefficient (common
for all datasets).

If not an instrumental effect or some long-term astrophysical variation, this trend can
be explained through a gravitational effect from a distant unseen companion. But such a
companion would also affect the observed transit times via the light-travel effect.
Basically, the light-travel effect is determined by the position of the inner star-planet
system along the line of sight, which equals to the integral of the RV trend. Therefore,
the transit time delay (TTD) corresponding to~(\ref{rvtrend}) would be:
\begin{eqnarray}
{\rm TTD} &=& \const + \frac{c_0}{\mathcal C} (t-t_0) + \frac{c_1}{\mathcal C} \frac{(t-t_0)^2}{2} \nonumber\\
 &=& \const + \frac{P c_0}{\mathcal C} (n-n_0) + \frac{P^2 c_1}{\mathcal C} \frac{(n-n_0)^2}{2},
\label{ttdtrend}
\end{eqnarray}
where $\mathcal C$ is the speed of light. This formula becomes mathematically identical to
the quadratic TTV trend~(\ref{trend}). But now the second term in~(\ref{ttdtrend}) with
$c_0$ (the absolute RV) represent a minor Doppler-like correction of the orbital period
$P$, while $c_1$ is basically the same TTV effect as observed e.g. for WASP-12.

Now, we simply have $T_{\rm d} = \mathcal C/c_1$. Taking into account the adopted
measurement units,
\begin{equation}
T_{\rm d} [{\rm Myr}] \simeq -\frac{300}{c_1 [{\rm m}\, {\rm s}^{-1}\, {\rm yr}^{-1}]}.
\label{ttdra}
\end{equation}
The minus sign appears because the positive RV trend means increasingly late light arrival
(apparent delay of the transits, as if the planet was spiraling out).

We found the statistically significant RV trend in the Corot-2, about $-74\pm 11$~m/s per
year. However, this trend can also be a hidden red-noise effect, because it disappears in
the alternative fit without splitting the HARPS data per subsets. One of the HARPS subsets
appeared to have just seven observations, so the ``split'' fit might appear rather
unreliable statistically. Finally, we found several RV outliers for Corot-2, possibly
indicating some hidden anomalies in its spectrum that suggest potential inaccuracies of the
derived Doppler information. Therefore, we conclude that the suspected radial acceleration
in Corot-2 remains controversial and needs further confirmation.

For Corot-2, the formula above implies $T_{\rm d} = (4.0\pm 0.6)$~Myr, or $q=T_{\rm d}^{-1}
\sim (0.25\pm 0.04)$~Myr$^{-1}$. The current TTV data are unable to reliably detect such a
trend. Finally, from the transit times we obtain $q\sim -0.1\pm 0.1$. The difference from
the RV estimate of $q$ is inconclusive given the high level of model dependency.
Nonetheless, Corot-2 remains an interesting target for further monitoring, because it may
provide the first detection of an unseen object simultaneously by Doppler and Roemer
effects.

Contrary to Corot-2, in WASP-12 we detect a more model-stable radial acceleration about
$-5$ or $-8$~m/s/yr, depending on the model. The trend information mainly comes from the
rich set of SOPHIE data available for this target ($>100$ points). The RV trend looks
rather reliable given its uncertainty, so the ``observed'' $T_{\rm d}$ need to be cleaned
from the non-tidal portion. This is performed in Sect.~\ref{sec_wasp12} below.

For WASP-4, the RV trend estimation is consistent with zero, but given the uncertainty it
may be as small as $-4$~m/s/yr (the three-sigma limit). This infers a TTV trend with
$T_{\rm d}\sim 70$~Myr, which is consistent with the observed transit times within two
sigma (see third panel in Fig.~\ref{fig_WASP4-lik}). Therefore, we cannot conclusively rule
out or confirm the tidal nature of the putative WASP-4 TTV trend, even if this trend
exists. More Doppler observations are needed to rule out the Roemer effect interpretation,
or to determine in which fraction this TTV trend can be explained so. Using the same method
as for WASP-12 (see Sect.~\ref{sec_wasp12}), we obtain for WASP-4 an estimate of the tidal
portion of the TTV trend of $q^{\rm tidal} = (0.019\pm 0.025)$~Myr$^{-1}$, again a
statistically insignificant value.

Yet another candidate with radial acceleration is WASP-3 ($-8$~m/s/yr), but it has just
$13$ RV measurements. Also, HAT-P-13 has an RV trend about $18$~m/s/yr, established long
ago \citep{Winn10}. Neither of these cases can be verified by TTV due to the lack of data.

Concerning the search of possible periods in the RV residuals, we leave this for a future
work. But here we confirm that Corot-2, HAT-P-13, WASP-4, and WASP-12 reveal no hints of
periodicity, so the corresponding RV trend estimations should not be affected by such
signals. Also, no residual variations were detected, including trends, in the updated RV
data for GJ~436 (though a spurious RV trend appeared in the older data release by
\citealt{Baluev15a}).

\section{Truly tidal portion in the WASP-12 TTV trend}
\label{sec_wasp12}
According to \citep{Bechter14}, WASP-12 is a member of a triple star system. It has a
stellar companion which appears as binary itself. The total mass of these two components B
and C is $0.75M_\odot$, and they may induce the radial acceleration of up to $0.33$~m/s/yr
on the primary star \citep{BaileyGoodman19}. Therefore, the radial acceleration of
$-7.5$~m/s/yr would definitely belong to some other unseen companions, possibly other
distant planets, or brown dwarfs, or even to an unresolved cool star.

Whatever object or multiple objects induced this radial acceleration, they should also
induce a quadratic TTV trend according to the formula~(\ref{ttdtrend}). Notice that even if
the RV trend was not significant at all, we should include it in the model in order to
determine a more realistic uncertainty of the tidal part of $T_{\rm d}$. Our RV trend
estimation implies the same sign as the observed cumulative value of $dP/dn$, so the tidal
portion is smaller than the total observed TTV. To determine how much it is smaller, and
how uncertain this value is, we perform a brief additional calculation.

First of all, we notice that information about the RV and TTV trends comes from
qualitatively different observations: either solely from RV data, or solely from transit
timings. Therefore, we may expect that these two quantities are practically uncorrelated
(even though some negligible correlation may appear because e.g. the planetary orbital
period $P$ is shared between the RV and transit data). Thanks to this property, the
corresponding trend coefficients may be treated independently from each other.

In terms of the TTV trend, its magnitude is defined by $q=T_{\rm d}^{-1}$, where $T_{\rm
d}$ from Table~\ref{tab_scp} is $(3.47\pm 0.24)$~Myr. Here we notice again that this
uncertainty is likely underestimated by the factor of $\sqrt{\chi^2_{\rm TTV}}$ from
Table~\ref{tab_traw}, or $1.33$. Therefore, a more realistic estimate is $T_{\rm d} =
(3.47\pm 0.32)$~Myr. The radial acceleration $c_1$ is estimated in Table~\ref{tab_scs}.
Based on the Corot-2 example, we believe that the alternative fit WASP-12' is more
realistic in terms of the RV trend, because in the basic fit (for split RV data) the
estimated radial acceleration may also include a local red noise effect appearing within
individual short in-transit runs. We do not expect that the estimated $c_1$ value may have
any significant escaped random uncertainty, since the stellar activity was already modelled
through the white and red RV jitter. Based on the formula~(\ref{ttdra}), from $c_1 =
(-7.5\pm 2.2)$~m/s/yr we obtain the apparent orbital decay time $T_{\rm d}^{\rm RA} = (40
\pm 12)$~Myr.

The tidal part of the TTV trend is then defined as:
\begin{equation}
\frac{1}{T_{\rm d}^{\rm tidal}} = \frac{1}{T_{\rm d}} - \frac{1}{T_{\rm d}^{\rm RA}}.
\end{equation}
Since $T_{\rm d}$ and $T_{\rm d}^{\rm RA}$ are practically uncorrelated, this formula
yields an estimation $T_{\rm d}^{\rm tidal} = (3.80\pm 0.40)$~Myr. Therefore, the truly
tidal orbital decay time is $\sim 10$ per cent larger than the observed (apparent) $T_{\rm
d}$, and also it has a larger uncertainty. Although the bias is only $\sim 1\sigma$, the
increase of the uncertainty is noticeable.

In this work we did not consider the alternative explanation of the WASP-12 TTV via the
apsidal precession \citep{Patra17}, although this model should also be corrected for the
nontidal part of the TTV.

\section{Conclusions and discussion}
The main conclusion of our work is that TTV data present in published literature are
significantly inhomogeneous. Therefore, they cannot be plainly processed by merging them
with each other. Substantial efforts must be made to increase the degree of homogeneity of
the TTV data by reprocessing the archival lightcurves, or by employing sophisticated noise
models separating different TTV subsets from each other. A lack of careful analysis of
heterogeneous TTV data may prove dangerous, potentially leading to spurious analysis
artefacts. In view of this it appears necessary to always verify our conclusions against
different data models, in order to gain an impression of how much they are model-dependent.

For the particular cases of WASP-12 and WASP-4 we notice that the Roemer effect induced by
possible additional companions may cause biased interpretation of the results. In
particular, the apparent value of the TTV trend may appear biased, and even mimicking tidal
phenomena. The Doppler observations are crucial in verifying such cases. In particular,
they may help to assess a more realistic uncertainty of the truly tidal part of a TTV
trend. Note that even if the RV trend estimation is consistent with zero it is important to
keep it as a free parameter, because its uncertainty increases the uncertainty in the tidal
portion of a TTV.

When performing the lightcurve analysis, the primary nuisance effect that currently remains
rather poorly modelled is the spot activity that implies anomalies in the transit curve. It
induces an additional TTV noise, which is difficult to predict or assess, except through
the post-hoc estimation of the TTV scatter. Various self-consistent (``global'') fits that
avoid using transit times as intermediate data do not include, as a rule, the spot-transit
effect in their uncertainties. To further improve their quality, we need to spend more
efforts to reducing the spot-transit event, e.g. to perform some kind of their automated
detection followed by a dedicated fitting or removing the associated piece of a lightcurve.
In our work the red noise model did not appear effective enough in removing the effect of
spot-transit anomalies, since from the statistical point of view such anomalies may behave
closer to a blue noise rather than red one.

We also consider the limb-darkening coefficients by \citet{ClaretBloemen11}, and conclude
that their FCM versions are rather accurate, though there is a detectable remaining bias in
the coefficient $B$ of about $0.1$. The LF version of these coefficients is rather poor for
practical use, as well as the older coefficients from
\citep{Claret00,Claret04}.

\section*{Acknowledgements}
Organization of the EXPANSION project (E.N.~Sokov, I.A.~Sokova), programming and
statistical analysis (R.V.~Baluev), and collection of literature data (V.Sh.~Shaidulin),
apart from the observations, were supported by the Russian Science Foundation grant
19-72-10023. Students of Horten Upper Secondary School (Horten Videreg\r{a}ende Skole)
H.~Fjeldstad, S.~Herstad,  D.N.~L{\o}nvik, I.~Oknes, and W.~Zhao are acknowledged for
taking part in observing WASP-12~b on the Nordic Optical Telescope. \"Ozg\"ur Ba\c{s}t\"urk
thanks The Scientific and Technological Research Council of Turkey (T\"UB\.{I}TAK) for
their support through the research grant 118F042 and for a partial support in using T100
telescope with the project No. 17BT100-1196. E.~Pak\v{s}tien\.e acknowledges support from
the Research Council of Lithuania (LMTLT) through grant 9.3.3-LMT-K-712-01-0103, and also
acknowledges observing time with $165$~cm and $35$~cm telescopes located at Mol\.etai
Astronomical Observatory in Lithuania. Administration of Ulugh Beg Astronomical Institute
(Tashkent, Uzbekistan), M.A. Ibrahimov, R.G. Karimov acknowledge financial support and
exchange visitor support from Max Planck Institute for Astronomy (Heidelberg, Germany). All
authors would like to thank the TRAPPIST team for sharing their archival data. Finally, we
thank the anonymous reviewer for their work and fruitful suggestions about the manuscript.




\bibliographystyle{mnras}
\bibliography{ttv}





\bsp	
\label{lastpage}
\end{document}